%% file: ANA-TRIG-2019-04-PAPER.tex
\newcommand*{\ATLASLATEXPATH}{}
\documentclass[cernpreprint,atlasdraft=true,texlive=2016, UKenglish]{\ATLASLATEXPATH atlasdoc}
 
\usepackage[subfigure]{\ATLASLATEXPATH atlaspackage}
\usepackage{\ATLASLATEXPATH atlasbiblatex}
 
\usepackage{\ATLASLATEXPATH atlasphysics}
 
\graphicspath{{logos/}{figures/}}
 
\usepackage{ANA-TRIG-2019-04-PAPER-defs}
 
\usepackage{multirow}

\AtlasTitle{Operation of the ATLAS Trigger System in Run 2}
 
\AtlasAbstract{
The ATLAS experiment at the Large Hadron Collider employs a two-level trigger system to record data at an average rate of 1 kHz from physics collisions, starting from an initial bunch crossing rate of 40~MHz.
During the LHC Run 2 (2015--2018), the ATLAS trigger system operated successfully with excellent performance and flexibility by adapting to the various run conditions encountered and has been vital for the ATLAS Run-2 physics programme.
For proton--proton running, approximately 1500 individual event selections were included in a trigger menu which specified the physics signatures and selection algorithms used for the data-taking, and the allocated event rate and bandwidth. The trigger menu must reflect the physics goals for a given data collection period, taking into account the instantaneous luminosity of the LHC and limitations from the ATLAS detector readout, online processing farm, and offline storage.
This document discusses the operation of the ATLAS trigger system during the nominal proton--proton data collection in Run 2 with examples of special data-taking runs. Aspects of software validation, evolution of the trigger selection algorithms during Run 2, monitoring of the trigger system and data quality as well as trigger configuration are presented.

}
 
\author{The ATLAS Collaboration}
 
\AtlasRefCode{TRIG-2019-04}

\PreprintIdNumber{CERN-EP-2020-109}
 
\AtlasJournalRef{JINST 15 (2020) 10, P10004}
\AtlasDOI{10.1088/1748-0221/15/10/P10004}
 
\hypersetup{pdftitle={Operation of the ATLAS trigger system in Run 2},pdfauthor={The ATLAS Collaboration}}
 
\begin{document}
 
\maketitle
 
\tableofcontents
 
\pagestyle{plain}
 
\section{Introduction}
 
The Trigger and Data Acquisition (TDAQ) system~\cite{ATLAS:2003aa} is an essential component of the ATLAS experiment~\cite{PERF-2007-01} at the Large Hadron Collider (LHC)~\cite{Evans:2008zzb} as it is responsible for deciding in real time whether to record data from a given collision.
Its successful operation has a crucial impact on the dataset used in physics analyses.
The TDAQ system operated efficiently during the LHC \runi (2009--2013)~\cite{TRIG-2011-02} at instantaneous luminosities up to $8\times\lumi{e33}$ and primarily at centre-of-mass energies of \SI{7}~\TeV\ and \SI{8}~\TeV, collecting more than three~billion proton--proton ($pp$) collision events.
 
In preparation for \runii (2015--2018), the TDAQ system underwent substantial upgrades and modifications to cope with the challenging conditions expected for data-taking during those years~\cite{TRIG-2011-02, TRIG-2016-01, Borodin:2015kfq}.
In \runii, the higher instantaneous luminosities, which surpassed the original design luminosity of $\lumi{e34}$, as well as the larger number of interactions per bunch crossing, and an increased centre-of-mass energy of \SI{13}~\TeV\ led to a significant increase in the event rate from interesting physics processes.
In addition to $pp$ collisions, the LHC provided  heavy-ion collisions at the end of each data-taking year. In 2015 and 2018, Pb+Pb collisions were recorded at a centre-of-mass energy per nucleon of \SI{5.02}~\TeV, in 2016 two datasets with $p$+Pb collisions were recorded at \SI{5.02}~\TeV\ and \SI{8.16}~\TeV, and in 2017 Xe+Xe collisions were recorded at \SI{5.44}~\TeV.
The operation of the trigger system under these challenging $pp$ data-taking conditions is described in this document.
 
This paper is organised as follows.
The ATLAS detector and TDAQ system is described in Sections~\ref{sec:ATLAS} and~\ref{sec:TDAQ}.
This is followed by a brief description of the LHC fill cycle, the fill patterns used in \runii and a description of a typical ATLAS run in Section~\ref{sec:AtlasTriggerOperation}.
Section~\ref{sec:shifter} introduces the `shifter and expert' model of the ATLAS trigger operations team.
The following two sections, Section~\ref{sec:TriggerMenu} and Section~\ref{sec:special}, discuss the trigger selections for the nominal $pp$ data-taking together with their evolution during \runii and provide a few examples of special runs, respectively.
Section~\ref{sec:conditions} presents changes of configuration and conditions that can be updated in real time during data-taking, while Section~\ref{sec:TriggerConfig} discusses the trigger configuration, database design and prescale mechanism.
The validation and sign-off cycle of the trigger software used during data-taking is described in Section~\ref{sec:ReleaseValidation}.
Section~\ref{sec:DebugStream} discusses the debug stream, which contains events for which no decision could be made.
Two sections present the monitoring options available during data-taking (Section~\ref{sec:OnlineMonitoring}), and the assessment of the quality of the fully reconstructed data (Section~\ref{sec:OfflineMonitoringDQ}). The conclusions are presented in Section~\ref{sec:conclusion}.

\section{ATLAS detector}
\label{sec:ATLAS}
 
The ATLAS detector~\cite{PERF-2007-01} covers nearly the entire solid angle around the collision point.\footnote{ATLAS uses a right-handed coordinate system with its origin at the nominal interaction point (IP)
in the centre of the detector and the \(z\)-axis along the beam pipe.
The \(x\)-axis points from the IP to the centre of the LHC ring, and the \(y\)-axis points upwards.
Cylindrical coordinates \((r,\phi)\) are used in the transverse plane, \(\phi\) being the azimuthal angle around the \(z\)-axis.
The pseudorapidity is defined in terms of the polar angle \(\theta\) as \(\eta = -\ln \tan(\theta/2)\).
Angular distance is measured in units of \(\Delta R \equiv \sqrt{(\Delta\eta)^{2} + (\Delta\phi)^{2}}\).}
It consists of an inner tracking detector surrounded by a thin superconducting solenoid, electromagnetic and hadronic calorimeters, and a muon spectrometer incorporating three large superconducting toroidal magnets.
 
The inner-detector system is immersed in a \SI{2}{\tesla} axial magnetic field and provides charged-particle tracking in the range \(|\eta| < 2.5\).
The high-granularity silicon pixel detector covers the vertex region and typically provides four measurements per track, the first hit being normally in the insertable B-layer (IBL) installed before Run~2~\cite{ATLAS-TDR-2010-19,PIX-2018-001}.
It is followed by the semiconductor tracker (SCT) which usually provides eight measurements per track.
These silicon detectors are complemented by the transition radiation tracker (TRT), which enables radially extended track reconstruction up to \(|\eta| = 2.0\).
The TRT also provides electron identification information based on the fraction of hits above a higher energy-deposit threshold corresponding to transition radiation.
 
The calorimeter system covers the pseudorapidity range \(|\eta| < 4.9\).
Within the region \(|\eta|< 3.2\), electromagnetic calorimetry is provided by barrel and endcap high-granularity lead/liquid-argon (LAr) calorimeters, with an additional thin LAr presampler covering \(|\eta| < 1.8\) to correct for energy loss in material upstream of the calorimeters.
Hadronic calorimetry is provided by the steel/scintillator-tile calorimeter, segmented into three barrel structures within \(|\eta| < 1.7\), and two copper/LAr hadronic endcap calorimeters.
The solid angle coverage is completed with forward copper/LAr and tungsten/LAr calorimeter modules optimised for electromagnetic and hadronic measurements respectively.
 
The muon spectrometer (MS) uses  separate trigger and high-precision tracking chambers to measure the deflection of muons in a magnetic field generated by the superconducting air-core toroids.
The field integral of the toroids ranges between \num{2.0} and \SI{6.0}{\tesla\metre} across most of the detector.
A set of precision tracking chambers covers the region \(|\eta| < 2.7\) with three layers of monitored drift tubes, complemented by cathode-strip chambers (CSCs) in the forward region where the background is highest.
The muon trigger system covers the range \(|\eta| < 2.4\) with resistive-plate chambers (RPCs) in the barrel, and thin-gap chambers (TGCs) in the endcap regions.
 
\section{ATLAS trigger and data acquisition system}
\label{sec:TDAQ}
 
\begin{figure}
\centering
\includegraphics[width=0.99\textwidth]{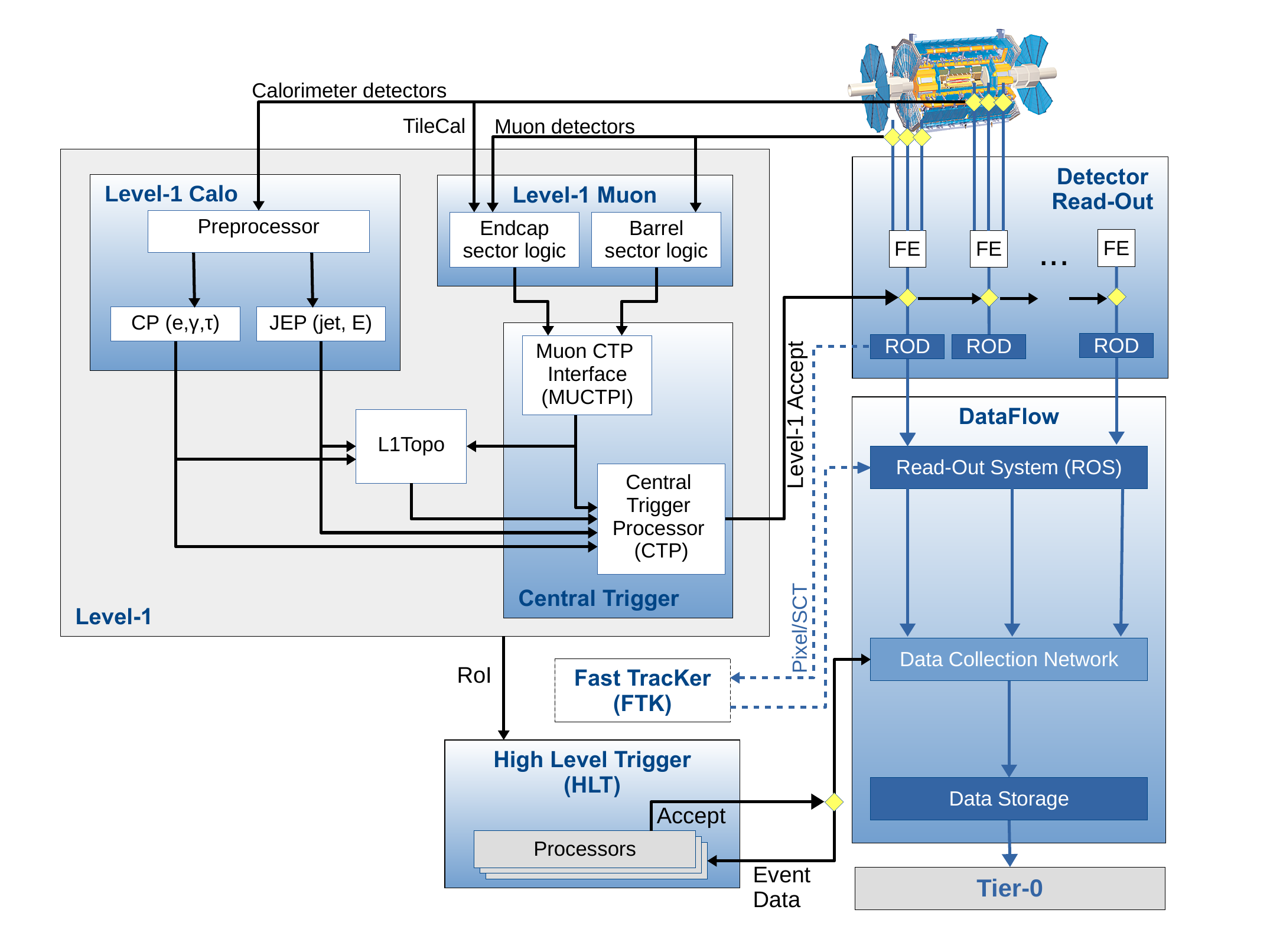}
\caption{The ATLAS TDAQ system in \runii showing the components relevant for triggering as well as the detector read-out and data flow.}
\label{fg:tdaqdiagram}
\end{figure}

The ATLAS TDAQ system is responsible for online processing, selecting and storing events of interest for offline analysis, and is shown diagrammatically in Figure~\ref{fg:tdaqdiagram}.
Events are selected using a two-stage trigger system~\cite{TRIG-2016-01}.
 
The Level-1 (L1) trigger is a hardware-based system that uses custom electronics to trigger on reduced-granularity information from the calorimeter and muon detectors~\cite{Aad:1602235}.
The L1 calorimeter (L1Calo) trigger takes signals from the calorimeter detectors as input~\cite{Achenbach_2008}.
The analogue detector signals are digitised and calibrated by the preprocessor, and sent in parallel to the Cluster Processor (CP) and Jet/Energy-sum Processor (JEP).
The CP system identifies electron, photon, and $\tau$-lepton candidates above a programmable threshold, and the JEP system identifies jet candidates and produces global sums of total and missing transverse energy.
The signals from the LAr calorimeter are bipolar and span multiple bunch crossings, which introduces a dependence of the amplitude on the number of collisions occurring in neighbouring bunch crossings (out-of-time pile-up).
Objects with narrow clusters such as electrons are not strongly affected by small shifts in energy, however the missing transverse momentum is very sensitive to small systematic shifts in energy over the entire calorimeter.
These effects are mitigated in the L1Calo trigger by a dedicated pedestal correction algorithm implemented in the firmware~\cite{collaboration2020performance}.
 
The L1 muon (L1Muon) trigger uses hits from the RPCs (in the barrel) and TGCs (in the endcaps) to determine the deviation of the hit pattern from that of a muon with infinite momentum~\cite{Aad:2020uyd}.
To reduce the rate in the endcap regions of particles not originating from the interaction point, the L1Muon trigger applies coincidence requirements between the outer and inner TGC stations, as well as between the TGCs and the tile calorimeter.
 
The L1 trigger decision is formed by the Central Trigger Processor (CTP), which receives inputs from the L1Calo trigger, the L1Muon trigger through the L1Muon Central Trigger Processor Interface (MUCTPI) and the L1 topological (L1Topo) trigger~\cite{Simoniello:2649959} as well as trigger signals from several detector subsystems such as the Minimum Bias Trigger Scintillators (MBTS)~\cite{MBTS}, the LUCID Cherenkov counter~\cite{LUCID} and the zero-degree calorimeter (ZDC)~\cite{ZDC}.
 
The CTP is also responsible for applying \textit{dead time}, which is a mechanism to limit the number of L1 accepts to be within constraints on detector read-out latency~\cite{Bertelsen:2063035}.
This preventive dead time limits the minimum time between two consecutive L1 accepts (simple dead time) to avoid overlapping read-out windows, and restricts the number of L1 accepts allowed in a given number of bunch crossings (complex dead time) to prevent front-end buffers from overflowing.
The complex dead time uses a leaky bucket model to emulate a front-end buffer. In this model, dead time is applied when the bucket is full. The size of the bucket is X, expressed in units of L1 accepts and R (in units of bunch crossings), the time it takes to read out one L1 accept. With these numbers the trigger rate, on average, is limited to X triggers in a time period of X~$\times$~R bunch crossings.
At the end of Run 2, the simple dead time setting was four bunch crossings, which corresponds to an inefficiency of about 1\% for a L1 rate of 90~kHz.
The complex dead time was configured with four different leaky bucket algorithms and one sliding-window algorithm to cover the read-out limitations of the various sub-detectors.
The total peak inefficiency was about 1\% for a L1 rate of 90~kHz.
 
The L1 trigger can select events by considering event-level quantities (e.g.~the total energy in the calorimeter), the multiplicity of objects above thresholds (e.g.~the transverse momentum of a muon, etc.), or by considering topological requirements (such as invariant masses or angular distances).
The topological requirements are applied in the L1Topo trigger to geometric or kinematic combinations between trigger objects received from the L1Calo or L1Muon systems.
The L1 trigger accepts events at a rate up to the maximum detector read-out rate of 100~kHz, down from the bunch crossing rate of about 40 MHz, within a latency of 2.5~$\mathrm{\mu}$s.
 
For each L1-accepted event, the Front-End (FE) detector electronics read out the event data for all detectors.
The data are sent first to ReadOut Drivers (RODs), performing the initial processing and formatting, and then to the ReadOut System (ROS) to buffer the data.
The data from the different sub-detectors are sent from the ROS to the second stage of the trigger, the High-Level Trigger (HLT), only when requested by the HLT.
In addition to performing the first selection step, the L1 triggers identify Regions-of-Interest (RoIs) in $\eta$ and $\phi$
within the detector to be investigated by the second trigger stage.

The second stage of the trigger, the HLT, is software-based.
A typical reconstruction sequence makes use of dedicated fast trigger algorithms to provide early rejection, followed by more precise and more CPU-intensive algorithms that are similar to those used for offline reconstruction to make the final selection.
These algorithms are executed on a dedicated computing farm of approximately 40\,000 selection applications known as Processing Units (PUs). Between each year of data taking, older hardware in the farm was replaced with newer hardware on a rolling basis to increase the available computing power and the total number of PUs.
The PUs are designed to make decisions within a few hundred milliseconds.
A step in such a sequence of algorithms will typically execute one or multiple feature-extraction algorithms requesting event-data fragments from within an RoI and terminate on a hypothesis algorithm which uses the reconstructed features to decide whether the trigger condition is satisfied or not.
In some cases, information from the full detector is requested in order to reconstruct physics objects (e.g.~for the reconstruction of the missing transverse momentum~\cite{Carlson:2686543}).
The HLT software is largely based on the offline software Athena~\cite{athena}, which itself is based on Gaudi~\cite{gaudi}, a framework for data processing for HEP experiments.
Gaudi/Athena is a component-based framework where each component (e.g.~algorithm, service, tool) is configured by a set of properties that can be defined during the configuration stage of the application.
The physics output rate of the HLT during an ATLAS data-taking run (see Section~\ref{ATLASrun}) is on average 1.2~kHz with an average physics throughput to permanent storage of 1.2~GB/s.
Once an event is accepted by the HLT, the Sub-Farm Output (SFO) sends the data to permanent storage for
offline reconstruction and exports the data to the Tier-0 facility~\cite{Tier0} at CERN's computing centre.
 
The Fast TracKer (FTK)~\cite{Shochet:1552953} is a hardware-based system for inner-detector track reconstruction designed to provide tracks to the HLT at the L1 accept rate. It was undergoing commissioning during \runii and was not used by the HLT for trigger decisions.

\section{\label{sec:AtlasTriggerOperation}LHC fill cycle, fill patterns and ATLAS runs}
 
In the following, the LHC fill cycle, the fill pattern and their representation in ATLAS, the so-called \textit{bunch groups}, are described.
Additionally, the ATLAS run, which refers to a continuous period of data acquisition that typically corresponds to an LHC fill cycle, is laid out.
 
\subsection{The LHC fill cycle}
\begin{figure}[tb]
\centering
\includegraphics[width=0.85\textwidth]{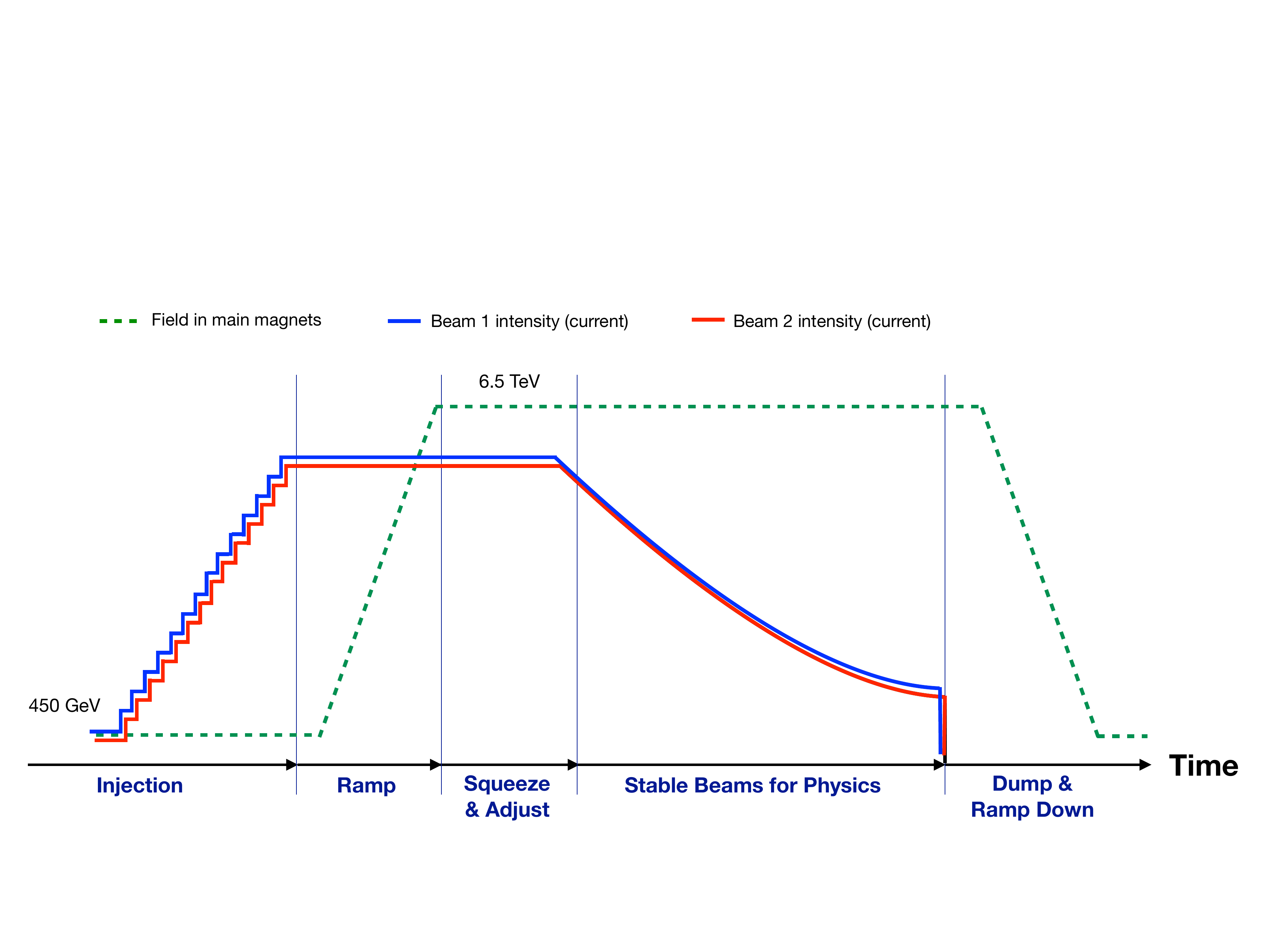}
\caption{The LHC goes through a cycle composed of several phases: the injection of beams into the rings, the acceleration to the collision energy during ramp, the preparation of beams for collisions during squeeze and adjust, the phase where collisions take place during stable beams, the extraction of the beams from the rings during dump, and finally the ramping down of the magnetic fields. Adapted from Ref.~\cite{LHCcycle}.}
\label{fig:LHCBeamModes}
\end{figure}

The LHC is a circular particle accelerator that is the last in a chain of accelerators used to bring particle beams into collisions at their final energies.
The beams travel through the LHC in opposite directions in separate rings of superconducting magnets, which are crossed at four interaction points.
The beams are kept separated in the interaction points using magnetic fields until they are ready for collisions.
The LHC aims to provide the largest, usable integrated luminosity of high-energy proton and ion collisions to the LHC experiments.
To provide collisions to the experiments, the LHC has to go through a cycle composed of several phases~\cite{LHCBeamModes}, which are shown in Figure~\ref{fig:LHCBeamModes}.
Each phase refers to one or several beam modes and all together are referred to as the \textit{nominal cycle}:
\begin{itemize}
\item \textit{Injection}: After the current in the magnets is increased to provide the field necessary for injection, 
beams are injected from the accelerator chain into the LHC rings following a \textit{filling scheme}, specifying the number of proton bunches and the spacing between them.
\item \textit{Ramp}: The beams are accelerated to the collision energy.
During this phase, the radio frequency systems accelerate the particles and the current in the magnets is further increased.
\item \textit{Squeeze} and \textit{adjust}: In these two phases, beams are prepared for collisions.
First, the beam sizes at the interaction points are reduced (\textit{squeeze}), then the beams are adjusted so that they are optimally colliding (\textit{adjust}).
\item \textit{Stable beams}: This is the phase when the LHC conditions are stable, collisions take place in the experiments, and it is safe for detectors to be turned on to record data.
Small adjustments of beam parameters are permitted~\cite{Hostettler:2300212,Karastathis:2648700}.
The LHC spent approximately 50\% of the time in stable beams throughout Run 2.
\item \textit{Dump} and \textit{ramp down}: Beams are extracted from the rings and safely dumped.
The dump can either be planned (by the LHC), requested (for example by experiments in case of problems with the detector) or unplanned.
Following the dump, the magnetic fields are ramped down.
\end{itemize}
The time in between two consecutive stable beams periods is referred to as \textit{turnaround}, and includes the nominal cycle as well as all necessary actions to set up the machine for operation with beams.
The ideal duration of the stable beams phase is typically 10--15~hours, depending on several factors, including luminosity lifetime, average turnaround duration, and predicted availability of the machine.

In about every second fill in the last year of \runii, fast luminosity scans were performed~\cite{PhysRevAccelBeams.21.102801} during stable beams to provide feedback on the transverse emittance at a bunch-by-bunch level to the LHC.
During these scans, beams are offset against each other in the $x$- and $y$-plane in several displacement steps.
The scans were typically done a few minutes after stable beams had been declared and just before the end of the stable beams period, and lasted a few minutes.

\subsection{LHC fill patterns in \runii}
During \runii, the LHC machine configuration evolved significantly. This was a major factor in improving luminosity performance each year of \runii~\cite{LHCRun2Config}.
The various bunch filling patterns used have a direct impact on the trigger configuration.
With the changing running conditions, adjustments had to be made in order to respect the trigger and DAQ system limitations (see e.g.\ Section~\ref{sec:lownumberbunches}).
 
At the start of \runii in 2015, the LHC used 50~ns bunch spacing and switched in August 2015 to the nominal 25~ns bunch spacing scheme~\cite{Bruning:782076}.
In June 2016, the high-brightness version of the 25~ns beam obtained through the Batch Compression, Merging and Splitting (BCMS) scheme~\cite{Damerau:1595719} became operational for physics production.
These changes brought about an increase in instantaneous luminosity to about $1.3\times\lumi{e34}$, resulting in higher trigger rates and an evolution in the trigger strategy.
In 2017, several LHC fills were dumped because of beam losses in the LHC sector 16L2~\cite{CERN-ACC-2019-012}. As a consequence, the 8b4e filling scheme~\cite{LHCInjPerf}  (eight bunches with protons, four bunches without protons) and later its high-brightness variant 8b4e BCS (Batch Compression and Splitting) with their low e-cloud build-up characteristics were made operational.
The 8b4e filling schemes circumvented the problem, but resulted in a reduction in the number of colliding bunches by 30\% compared to the BCMS scheme.
To compensate for the loss in luminosity due to the decrease in colliding bunches, the bunch intensity was increased and this led to a 33\% increase of simultaneous interactions per bunch crossing (`pile-up'), up to 80 interactions compared to up to 60 interactions previously.
Such an increase together with the high luminosities of up to $1.9\times\lumi{e34}$~\cite{ATLAS-CONF-2019-021} would have resulted in an increase of trigger rates, straining the CPU resources of the HLT farm.
The trigger configuration intended to be used for a luminosity of up to $2.0\times\lumi{e34}$ with a pile-up of 60 would have required higher trigger thresholds at a pile-up of 80, leading to a reduced efficiency for many physics analyses.
Therefore, ATLAS requested that the luminosity be kept constant for the first few hours of a run (\textit{luminosity-levelling}) at a luminosity of $1.56\times\lumi{e34}$ with a pile-up of 60.
In 2018, the LHC switched back to the 25~ns BCMS beam for luminosity production, as the problems with beam losses in 16L2 were mitigated~\cite{16L2fix}, and the pile-up interactions were again reduced to about 60.
More information about these filling schemes can be found in Ref.~\cite{LHCInjPerf}.
 
\subsection{\label{sec:BunchGroups}Bunch groups}
In the LHC, there are a total of 3564 bunch crossings per LHC revolution.
Each of these bunch crossings can have either two bunches colliding, one bunch, or be empty of protons.
Each bunch crossing is identified by a Bunch Crossing Identifier (BCID) from 0 to 3563.
A list of BCIDs is called a \textit{bunch group}.
 
\begin{figure}
\centering
\includegraphics[width=0.99\textwidth]{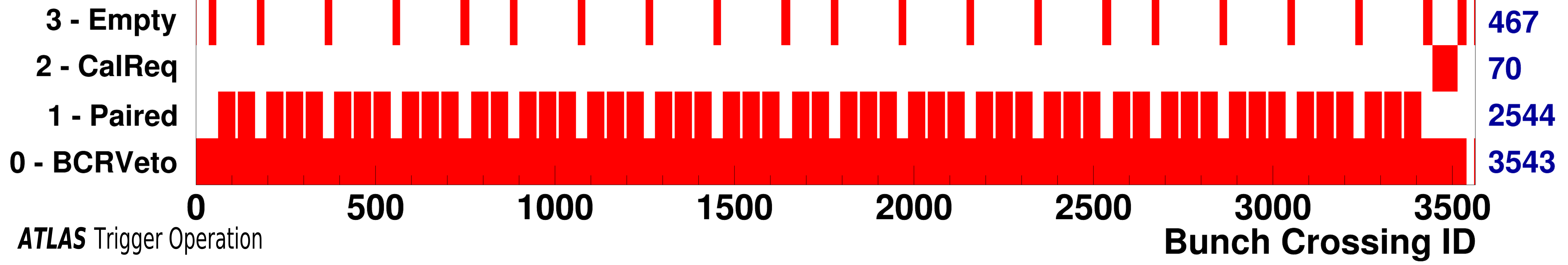}
\caption{Example bunch group configurations for four out of the 16 possible bunch groups. The numbers in blue on the right indicate the number of bunch crossings for each group. The group of the bunch counter reset veto (\textit{BCRVeto}) leaves a short time slice for distribution of the LHC bunch counter reset signal to the on-detector electronics. The \textit{Paired} bunch group indicates the bunch crossing IDs with colliding bunches, while the \textit{Empty} bunch group contains no proton bunches and is generally used for cosmic ray, noise and calibration triggers. The calibration requests (\textit{CalReq}) bunch group can be used to request calibration triggers.}
\label{fig:bunchgroup}
\end{figure}

Bunch group conditions are used in combinatorial logic `AND' with other trigger conditions to define which items generate a L1 accept.
There are 16 distinct bunch groups that can be defined in ATLAS, each with its own particular purpose, defined for each LHC bunch.
Figure~\ref{fig:bunchgroup} shows four types of bunch groups which are described in the following.
 
Bunch group conditions can be paired (colliding) bunches for physics triggers, single (one-beam) bunches for background triggers and empty bunches for cosmic ray, noise and calibration triggers.
More complex schemes are possible, e.g.\ requiring unpaired bunches separated by at least 75~ns from any bunch in the other beam.
Two bunch groups have a more technical purpose:
the calibration requests group defines the times at which sub-detectors may request calibration triggers, typically in the long gap with no collisions and the group of the bunch counter reset veto leaves a short time slice for distribution of the LHC bunch count reset signal to the on-detector electronics.

As the LHC filling scheme can vary from fill to fill, ATLAS has developed and commissioned a procedure for monitoring and redefining the bunch groups using dedicated electrostatic detectors.
These so-called \textit{beam pick-ups}~\cite{Ohm:1172170} are located 175~m upstream of the interaction point.
An online application measures the filling scheme seen by the beam pick-ups and calculates the corresponding bunch groups.
While the configuration of some bunch groups is given by the LHC (e.g.~the colliding BCIDs) through the fill pattern, others can be defined to contain any desired list of BCIDs for specific data-taking requests (e.g.~in van der Meer scans~\cite{ATLAS-CONF-2019-021} or single-beam background studies).
 
The 16 bunch group configurations are together called a \textit{bunch group set}, which is different for each different LHC filling scheme.
The bunch group sets are generated for each filling scheme in advance of running, using information about the positions of bunches with protons in each beam provided by the LHC.
The generated bunch group set is then checked against the measured beam positions to ensure that it matches.
The CTP pairs each L1 trigger item with a specific bunch group defined in the set.
Those L1 trigger items which are employed to select events for physics analyses trigger on bunch groups containing all colliding BCIDs.
The CTP can also provide random triggers and apply specific bunch crossing requirements to those. 
 
\subsection{\label{ATLASrun}The ATLAS run structure}
An ATLAS \textit{run} is a period of data acquisition with stable detector configuration and in the case of physics data-taking usually coincides with an LHC fill, which can last many hours.
Another example is a cosmic-ray data-taking run, which takes place when there is no beam in the LHC and the ATLAS detector is used to detect cosmic rays to study detector performance~\cite{PERF-2010-02}.
The DAQ system assigns a unique number to every run at its beginning.
An identifier is assigned to each event and is unique within the run (starting at 0 for each run).
A run is divided into Luminosity Blocks (LB), with a length of the order of one minute and identified by an integer unique within a given run.
A LB defines an interval of constant luminosity and stable detector conditions (including the trigger system and its configuration).
To define a data sample for physics, quality criteria are applied to select LBs where conditions are acceptable.
The instantaneous luminosity in a given LB is multiplied by the LB duration to obtain the integrated luminosity delivered in that LB.
The length of the LB can be changed during the run and a new LB can be started at any time (following a 10 second minimum delay).
From a data quality point of view, the LB represents the smallest quantity of data that can be declared good or bad for physics analysis.
 
\begin{figure}[tb]
\centering
\includegraphics[width=0.3\textwidth]{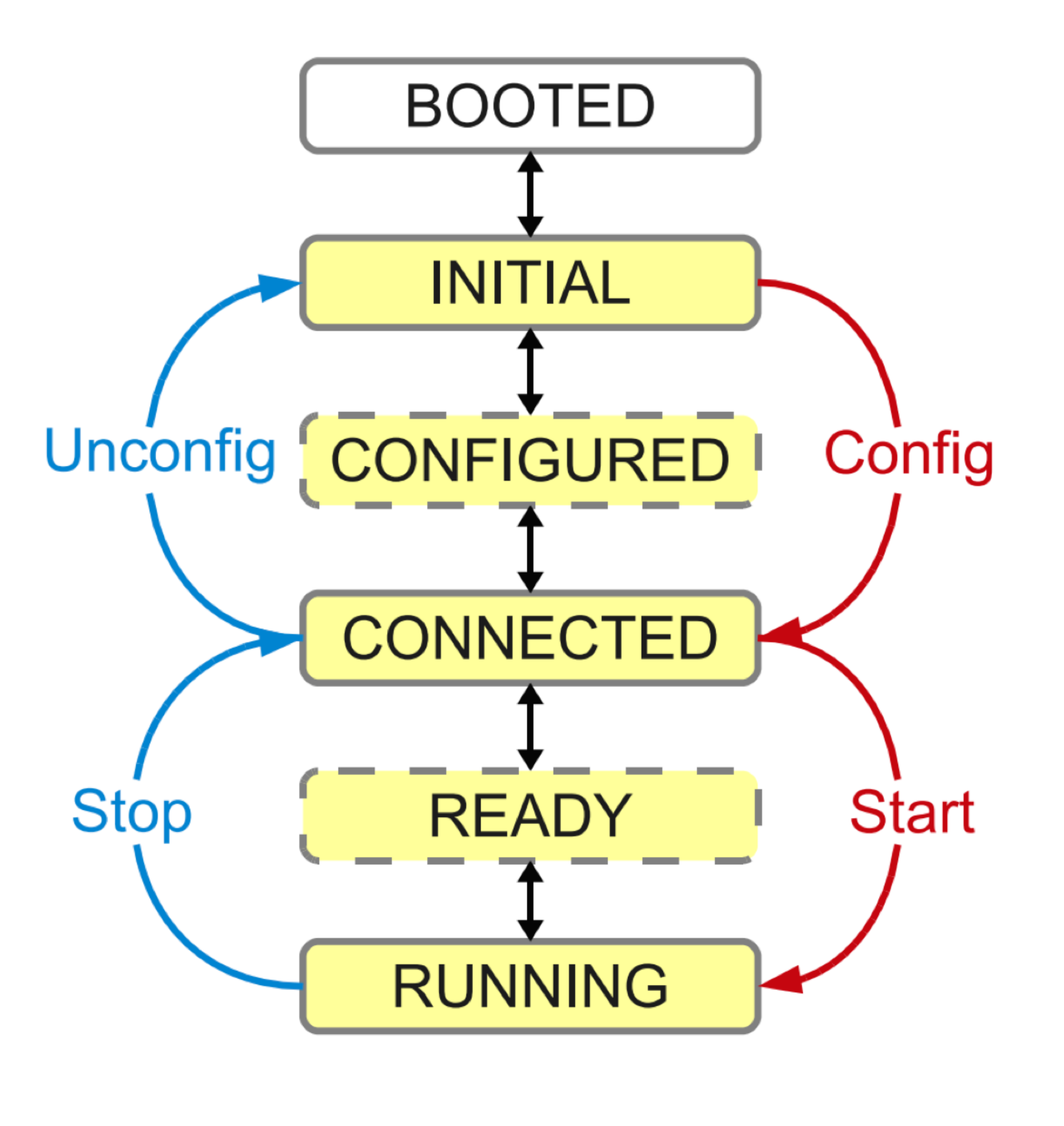}
\caption{The software and hardware components of the ATLAS detector follow the transitions of a finite-state machine used to synchronise the configuration of all applications and detectors within ATLAS.
Run-dependent configurations (e.g.~loading of conditions data) are performed during the start transition, which can take several minutes for the entire ATLAS detector~\cite{Winklmeier:1271370}.}
\label{fig:ATLASrunstates}
\end{figure}

To start a run, the software and hardware components of the ATLAS detector have to follow the transitions of a finite-state machine~\cite{Winklmeier:1271370}.
The transitions performed by the applications are shown in Figure~\ref{fig:ATLASrunstates}.
During `boot', all applications are being started.
In the `configure' and `connect' transitions, the hardware and applications are configured and connections between different applications are established where necessary.
Finally, during `start', a run number is assigned and the applications perform their final (run-dependent) configuration.
Once all applications arrived at the `ready state', the CTP releases the inhibit and events start flowing through the system.
 
\section{\label{sec:shifter}Operational model of the ATLAS trigger system}
During the operation of the LHC, the ATLAS detector is operated and monitored by a shift crew in the ATLAS control room (ACR), 24 hours a day, 7 days a week, supported by a pool of remote on-call experts.
Shifters and experts are responsible for the efficient collection of high-quality data.

The operation and data quality monitoring of the trigger system is overseen by two operation coordinators whose main responsibility is to ensure smooth and efficient data-taking.
They coordinate a team of weekly on-call experts, on rotation, for the areas listed below.
Operation coordinators and on-call experts work together closely at a daily trigger operation meeting to plan the activities of the day.
\begin{itemize}
\item \emph{ACR trigger desk}: during the shift in the ACR, the person is responsible for providing the needed trigger configuration and for monitoring the operation of the trigger system in close communication with other ACR shifters.
\item \emph{Online}: responsible for the proper operation of the ATLAS trigger and primary support for the ACR trigger shifter.
\item \emph{Trigger menu}: responsible for the preparation of the trigger configuration of active triggers and their prescale factors (see Section~\ref{sec:TriggerMenu}).
\item \emph{Online release}: collection and review of software changes and monitoring the state of the software release for online usage via validation tests that run every night; deployment of the online software release on the machines used during data-taking.
\item \emph{Reprocessing}: in charge of validating the online software release (see Section~\ref{sec:ReleaseValidation}) by running the simulation of the L1 hardware and the HLT software on a dedicated dataset to spot errors by running on large samples.
\item \emph{Data quality and debug stream}: responsible for the data quality assessment of recorded data; investigate and recover the events in the debug stream (see Section~\ref{sec:DebugStream}).
\item \emph{Signature-specific}: monitor the performance of triggers for signatures, assist in data quality assessment and reprocessing sign-off; several trigger signatures are grouped together (muon and $B$-physics and Light States; jet, missing transverse momentum and calorimeter energy clusters; $\tau$-lepton, electron and photon; $b$-jet signature and tracks).
\item \emph{Level-1}: each L1 trigger system (L1Calo, L1Muon barrel, L1Muon endcap, and CTP) has an on-call expert who helps to ensure smooth operation of the L1 trigger and monitors the data quality for their respective system.
\end{itemize}

In additon to data-taking, the trigger operation group participates in special runs of a technical nature together with the ATLAS DAQ team to develop and test the online software and tools to be used for data-taking.
It also provides support for other ATLAS systems during detector commissioning runs and for special tests during LHC downtime periods.
 
\pagestyle{plain}
 
\section{\label{sec:TriggerMenu}The Run-2 trigger menu and streaming model}
 
Events are selected by trigger chains, where a chain consists of a L1 trigger item and a series of HLT algorithms that reconstruct physics objects and apply kinematic selections to them.
Each chain is designed to select a particular physics signature such as the presence of leptons, photons, jets, missing transverse momentum, total energy and $B$-meson candidates.
The list of trigger chains used for data-taking is known as a trigger menu, which also includes prescales for each trigger chain.
To control the rate of accepted events, a prescale value, or simply prescale, can be applied.
For a prescale value of $n$, an event has a probability of 1/$n$ to be kept.
Individual prescale factors can be given to each chain at L1 or at the HLT, and can be any value greater than or equal to one.
More details of how prescales are applied can be found in Section~\ref{sec:prescales}.
 
The complete set of trigger selections must respect all trigger limitations and make optimal use of the available resources at L1 and the HLT (e.g., maximum detector read-out rate, available processing resources of the HLT farm, and maximum sustainable rate of permanent storage).
Rates and resource usage are determined as described in Section~\ref{sec:costmon} and Section~\ref{sec:ratePred}.
 
The configuration is driven by the physics priorities of the experiment, including the number of clients satisfied by a particular trigger chain.
The main goal of the Run-2 trigger menu design was to maintain the unprescaled single-electron and single-muon trigger \pT thresholds around 25 GeV despite the expected higher trigger rates to ensure the collection of the majority of events with leptonic $W$ and $Z$ boson decays.
The primary triggers (used for physics analyses and unprescaled)
cover all signatures relevant to the ATLAS physics programme including electrons, photons, muons, $\tau$-leptons, jets, $b$-jets and $E_\mathrm{T}$ which are used for Standard Model precision measurements including decays of the Higgs, $W$ and $Z$ bosons, and searches for physics beyond the Standard Model such as heavy particles, supersymmetry or exotic particles. A set of low transverse momentum dimuon triggers is used to collect $B$-meson decays, which are essential for the $B$-physics programme of ATLAS.
 
Heavy-ion (HI) collisions differ significantly from $pp$ collisions, and therefore require a dedicated trigger menu to record the data. The main components of the HI trigger menu are triggers selecting hard processes (high $E_\mathrm{T}$, $b$-jets, muons, electrons, and photons) in inelastic Pb+Pb collisions, minimum-bias triggers for peripheral and central collisions, triggers selecting events with particular global properties (event-shape triggers to collect events with large initial spatial asymmetry of the collisions, ultra-central collision triggers), as well as triggers selecting various signatures in ultra-peripheral collisions. More information about the HI trigger menu and associated streams can be found in Ref.~\cite{ATL-DAQ-PUB-2018-002}.
 
Apart from the trigger menu used to record nominal $pp$ or HI collisions, additional trigger menus were designed in Run 2 for special data-taking configurations, with some examples discussed in Section~\ref{sec:special}.

\subsection{The trigger menu evolution in \runii}
The trigger menu for $pp$ data-taking evolved throughout \runii due to the increase of the instantaneous luminosity and the number of pile-up interactions.
The composition of the trigger menu is developed based on the expected luminosity for each year, with looser selections deployed during early data-taking or when the peak luminosity falls below the predicted target value.
The main trigger chains that comprise the ATLAS trigger menu for 2015 targeting an instantaneous luminosity of $5\times\lumi{e33}$ and valid for a peak luminosity up to $6.5\times\lumi{e33}$ are described in detail along with their performance in Ref.~\cite{ATL-DAQ-PUB-2016-001}.
As the instantaneous luminosity increased substantially in 2016 (up to $1.3\times\lumi{e34}$) and again in 2017 (up to $1.6\times\lumi{e34}$), it became necessary to adjust the trigger menu each year accordingly.
The various improvements and the performance for the trigger menu used in 2016 and 2017 are described in detail in Refs.~\cite{ATL-DAQ-PUB-2017-001} and~\cite{ATL-DAQ-PUB-2018-002}, respectively.
The peak luminosity in 2018 was close to $2.0\times\lumi{e34}$.
Even though the luminosity was higher than in 2017, the number of interactions per bunch crossing was similar.
The resources needed to continue running the same trigger menu as in 2017 were estimated to fall within the limitations of the trigger system. The 2018 trigger menu~\cite{ATL-DAQ-PUB-2019-001} therefore only contained additions on top of the 2017 menu together with a few changes and improvements to the trigger selections used in 2017.
 
\subsection{\label{sec:costmon}Cost monitoring framework}
\begin{figure}
\centering
\includegraphics[width=0.48\textwidth]{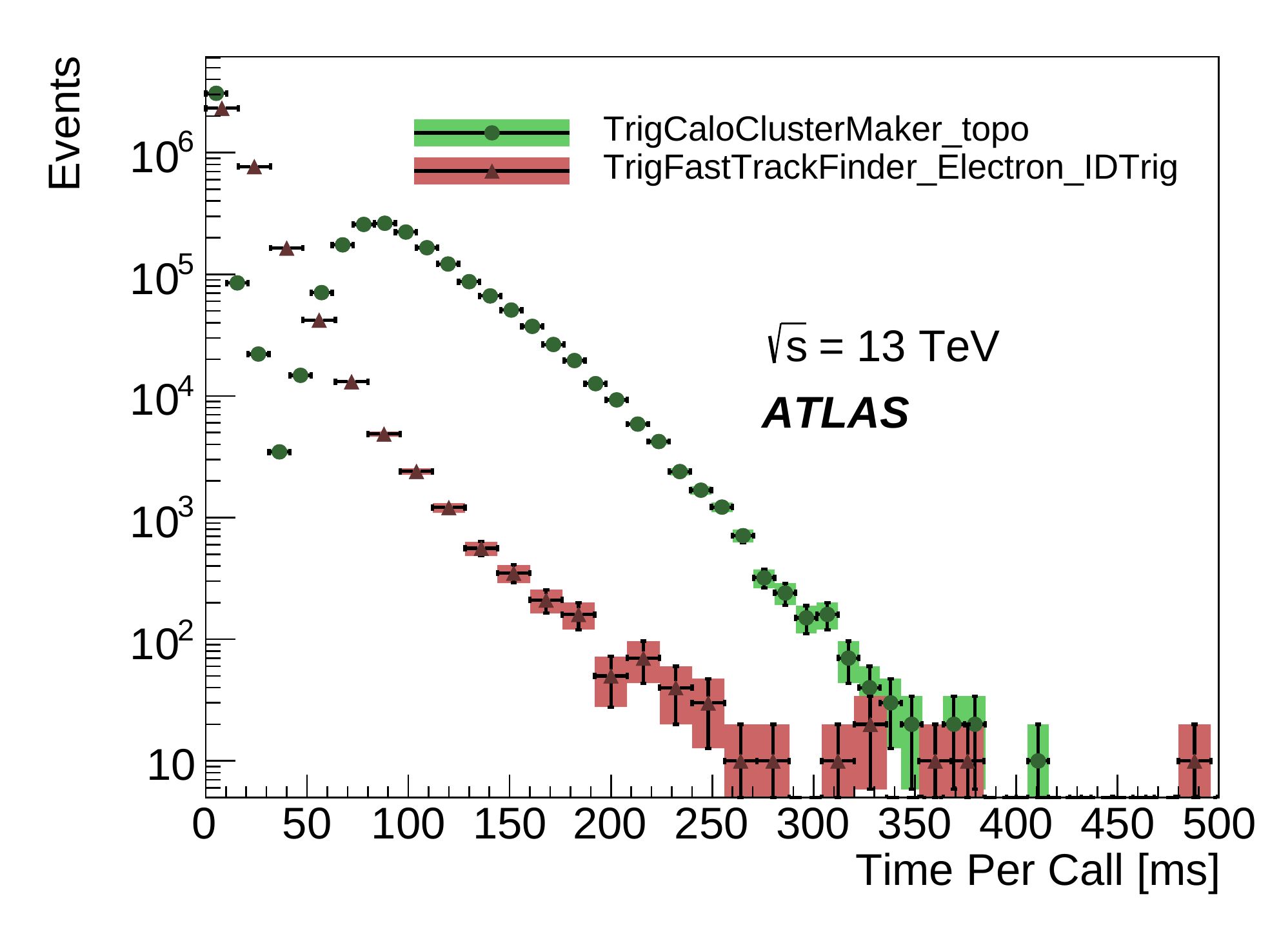}
\includegraphics[width=0.48\textwidth]{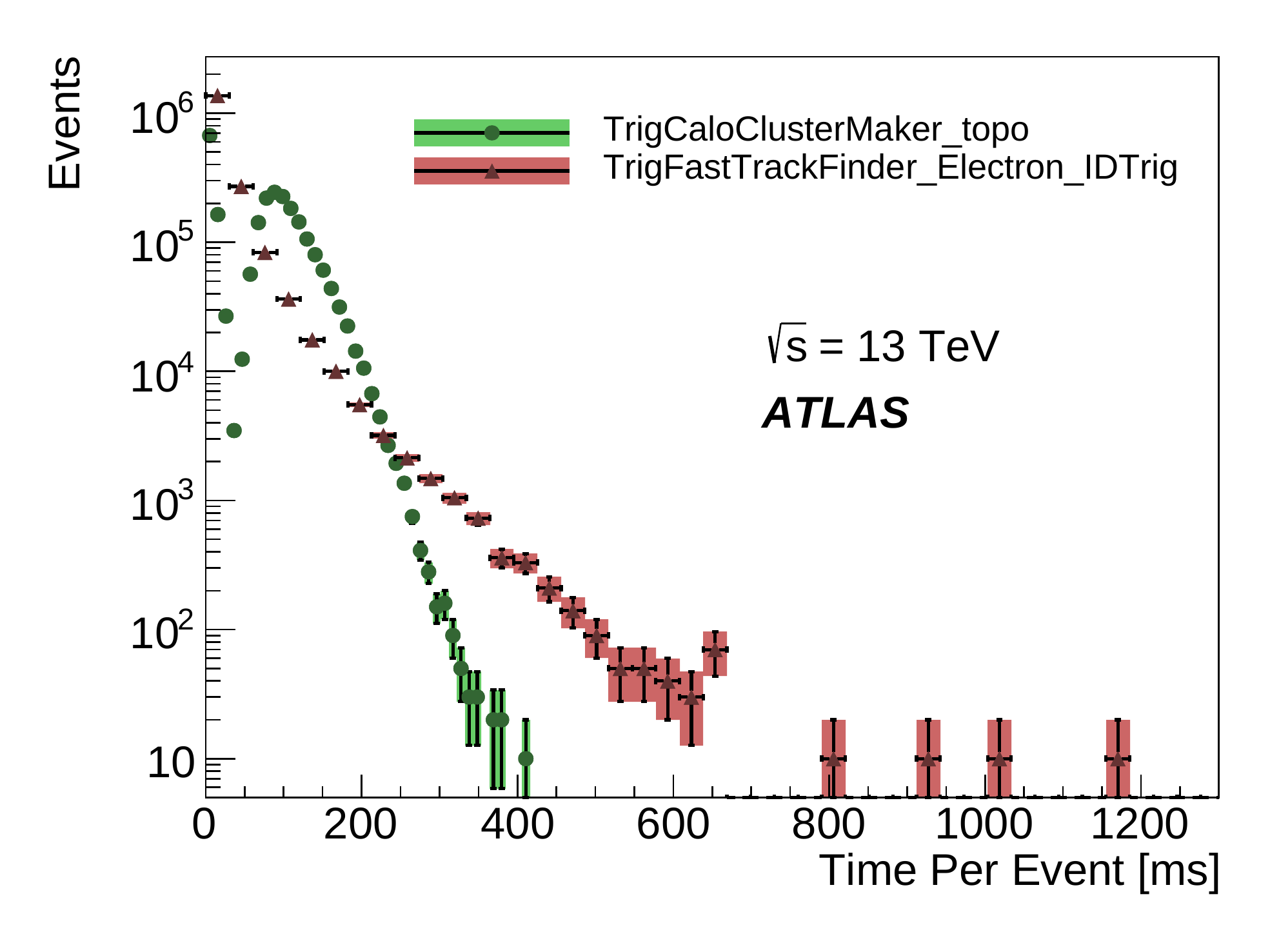}\\
\includegraphics[width=0.48\textwidth]{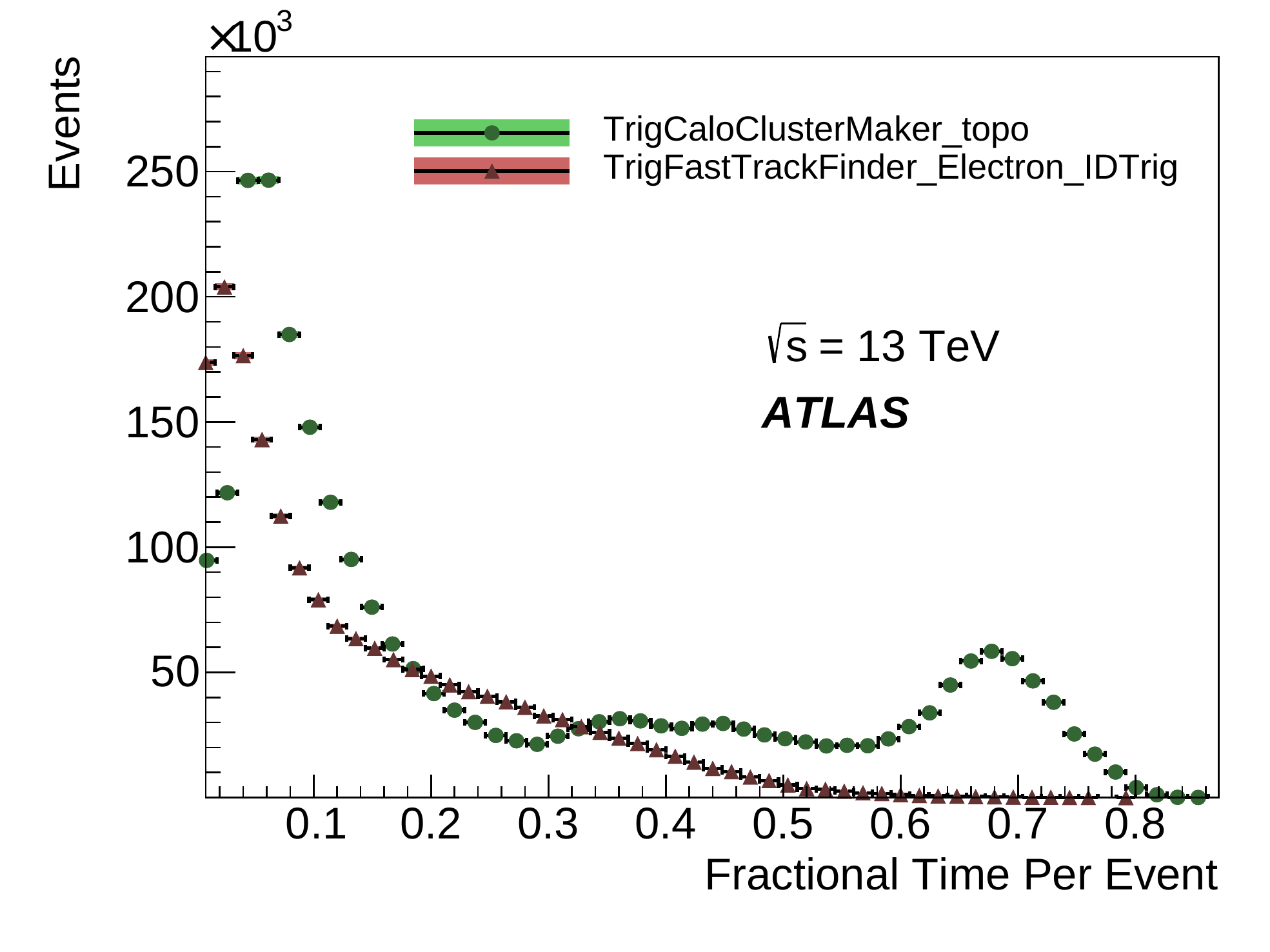}
\includegraphics[width=0.48\textwidth]{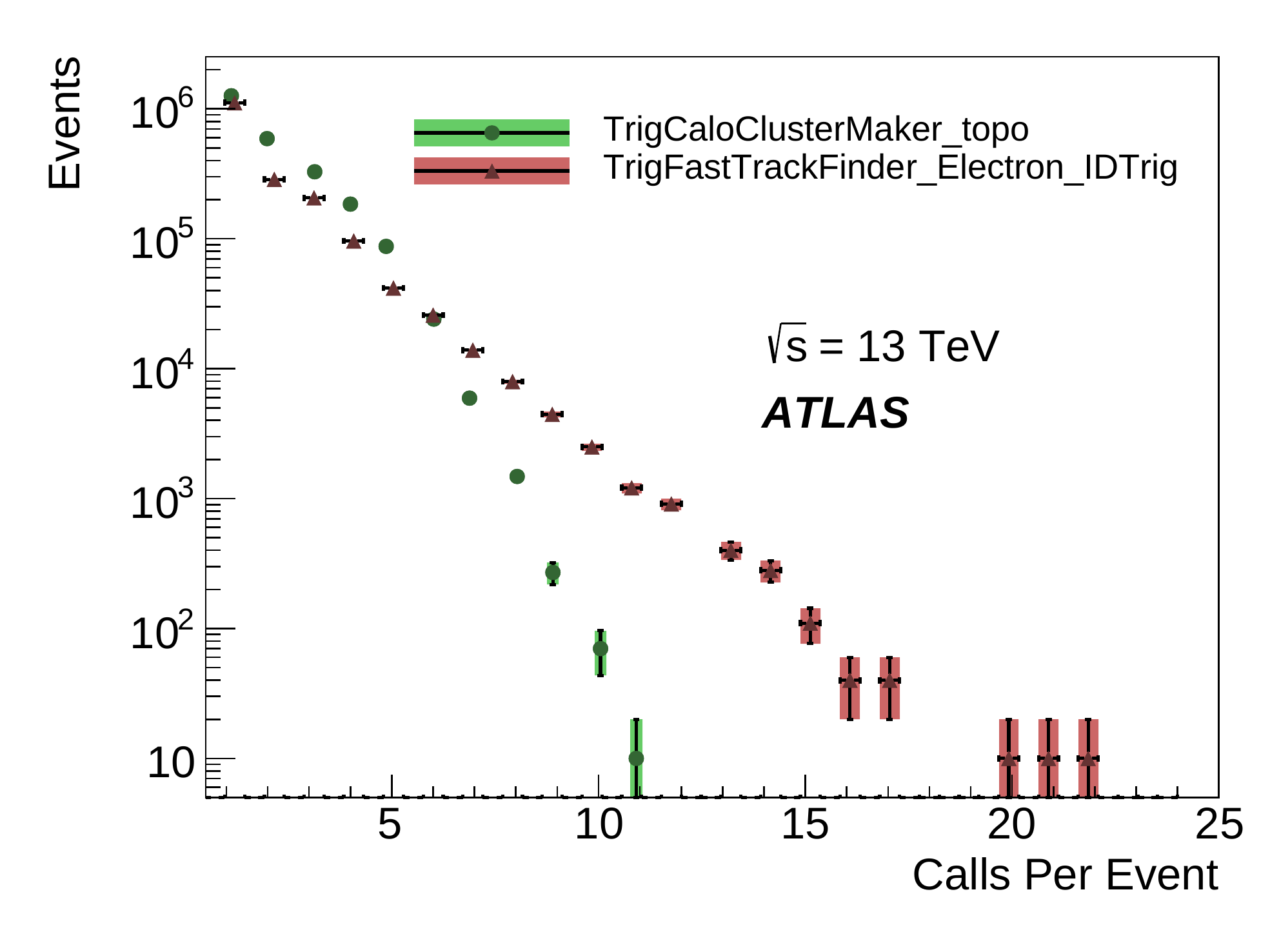}
\caption{Cost monitoring distributions for two HLT algorithms~\cite{Martin:2237426}: the topological clustering of calorimeter data ($\mathrm{TrigCaloCluserMaker\_topo}$) is shown in green and the inner-detector electron track identification ($\mathrm{TrigFastTrackFinder\_Electron\_IDTrig}$) is shown in red. Presented are the execution time (top) per call (left) and per event (right), as well as the execution time expressed as a fraction of the total execution time of all algorithms (bottom) in the event (left) and number of executions per event (right). Only statistical uncertainties are shown.}
\label{fig:costmon}
\end{figure}

The ATLAS cost monitoring framework~\cite{Martin:2237426} consists of a suite of tools to collect monitoring data on both CPU usage and data-flow over the data-acquisition network during the trigger execution.
These tools are executed on a sample of events processed by the HLT, irrespective of whether the events pass or fail the HLT selection.
It is primarily used to prepare the trigger menu for physics data-taking through the detailed monitoring of the system, allowing data-driven predictions to be made utilising dedicated datasets (enhanced bias dataset, see Section~\ref{sec:ratePred}).
Monitored data include algorithm execution time, data request size, and the logical flow of the trigger execution for all L1-accepted events.
To sample a representative subset of all L1-accepted events, a monitoring fraction of 10\% is chosen.
Example monitoring distributions are given for two of the many algorithms in Figure~\ref{fig:costmon}: calorimeter topological clustering~\cite{PERF-2014-07} and electron tracking.
These monitoring data were collected over a 180~s data-taking period at $1\times\lumi{e34}$.
Topological clustering can run either within an RoI or as a full detector scan, leading to a double-peak structure in the processing time as shown in Figure~\ref{fig:costmon} (top left).
Equivalently to the procedure of predicting the rates of individual HLT chains and trigger menus (see Sections~\ref{sec:ratePred} and~\ref{sec:ReleaseValidation}), it is possible to estimate the number of HLT PUs which will be required to run a given trigger chain or menu. This functionality was extremely useful in planning for different LHC scenarios in 2017 and in preparation for 2018 data-taking.
 
\subsection{Run-2 streaming model}
The trigger menu defines the streams to which an event is written, depending on the trigger chains that accepted the event.
Data streams are subdivided into files for each luminosity block, which facilitates the subsequent efficiency and calibration measurements under varying running conditions.
 
The five different types of data streams considered in the recording rate budget available at the HLT during nominal $pp$ data-taking are:
\begin{itemize}
\item Physics stream: contains events with collision data of interest for physics studies. The events contain full detector information and dominate in terms of processing, bandwidth and storage requirements.
\item Express stream: very small subset of the physics stream events reconstructed offline in real time for prompt monitoring and data quality checks.
\item Debug streams: events for which no trigger decision could be made are written to this stream. These events need to be analysed and recovered separately to identify and fix possible problems in the TDAQ system (see Section~\ref{sec:DebugStream}).
\item Calibration streams: events which are triggered by algorithms that focus on specific sub-detectors or HLT features are recorded in these types of streams. Depending on the purpose of the stream, only partial detector information is recorded through a strategy called Partial Event Building (PEB)~\cite{TRIG-2016-01}, which has the potential to significantly reduce the event size.
\item Trigger-Level Analysis (TLA) streams: events sent to this stream store only partial detector information and specific physics objects reconstructed by the HLT to be used directly in a physics analysis.
\item Monitoring streams: events are sent to dedicated monitoring nodes to be analysed online for, e.g., detector monitoring, but are not recorded.
\end{itemize}
 
For special data-taking configurations it is possible to introduce additional streams; an example is the recording of enhanced bias data, which is discussed in Section~\ref{sec:ratePred}.
With the exception of the debug streams, the streaming model is inclusive, which means that an event can be written to multiple streams.
Aside from the express stream, there are typically multiple different streams of each type.
For PEB, data are only stored for specific sub-detectors, or for specific regional fragments from specific sub-detectors.
Similarly, the TLA stream (see Ref.~\cite{ATLAS-CONF-2016-030} for more details of the procedures) only stores physics objects reconstructed by the HLT with limited event information and uses these trigger-level objects directly in a physics analysis~\cite{EXOT-2016-20}.
By writing out only a fraction of the full detector data, the event size is reduced, making it possible to operate these triggers at higher accept rates while not being limited by constraints on the output bandwidth. This strategy is effective in avoiding high prescales at the HLT for low transverse momentum (\pT) triggers.
 
\begin{figure}
\begin{center}
\includegraphics[width=\textwidth]{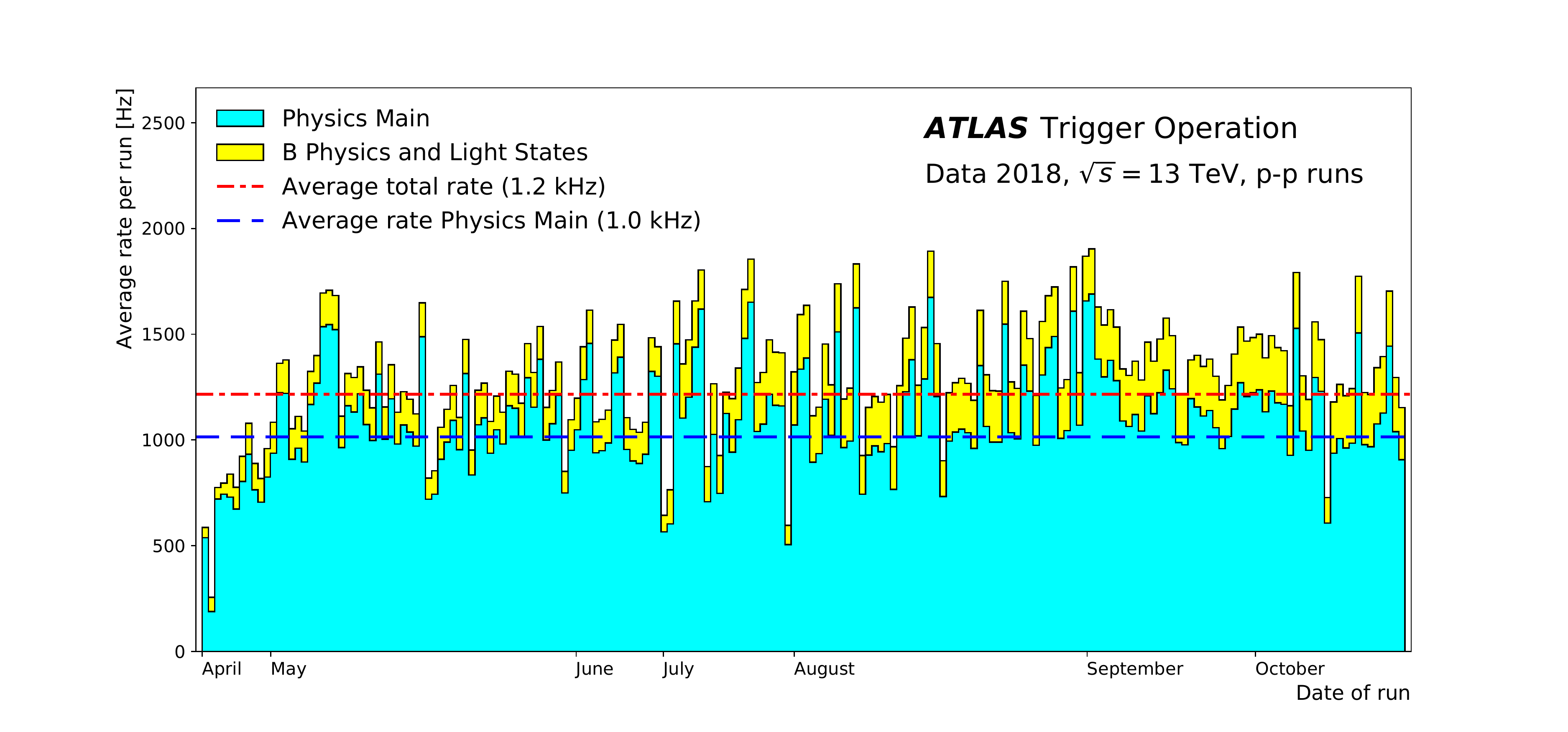}
\caption{The average recording rate of the main physics data stream and the BLS data stream for each ATLAS $pp$ physics run taken in 2018.
The average of all runs for these two streams is indicated as a red dash-dotted line, and the average of the main physics stream is indicated as a blue dashed line.}
\label{fg:streamRates}
\end{center}
\end{figure}

Figure~\ref{fg:streamRates} shows the average recording rate of the physics data streams of all ATLAS $pp$ runs taken in 2018.
Events for physics analyses are recorded at an average rate of $\sim$1.2~kHz. This comprises two streams, one dedicated to $B$-physics and Light States (BLS) physics data, which averaged 200~Hz, and one for all other main physics data, which averaged the targeted 1~kHz.
The BLS data are kept separate so the offline reconstruction can be delayed if available resources for processing are scarce.
 
\begin{figure}
\begin{center}
\includegraphics[width=0.7\textwidth]{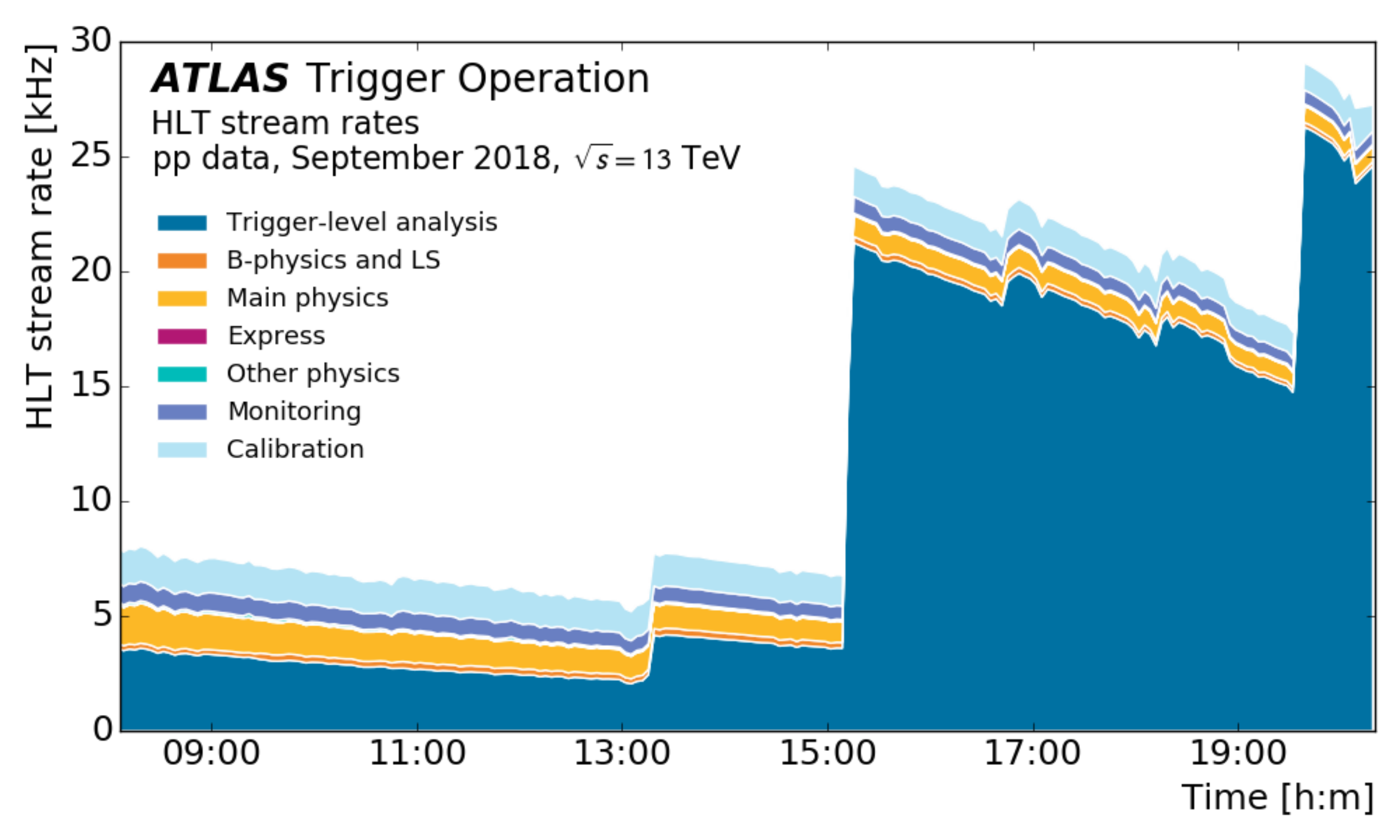}
\includegraphics[width=0.7\textwidth]{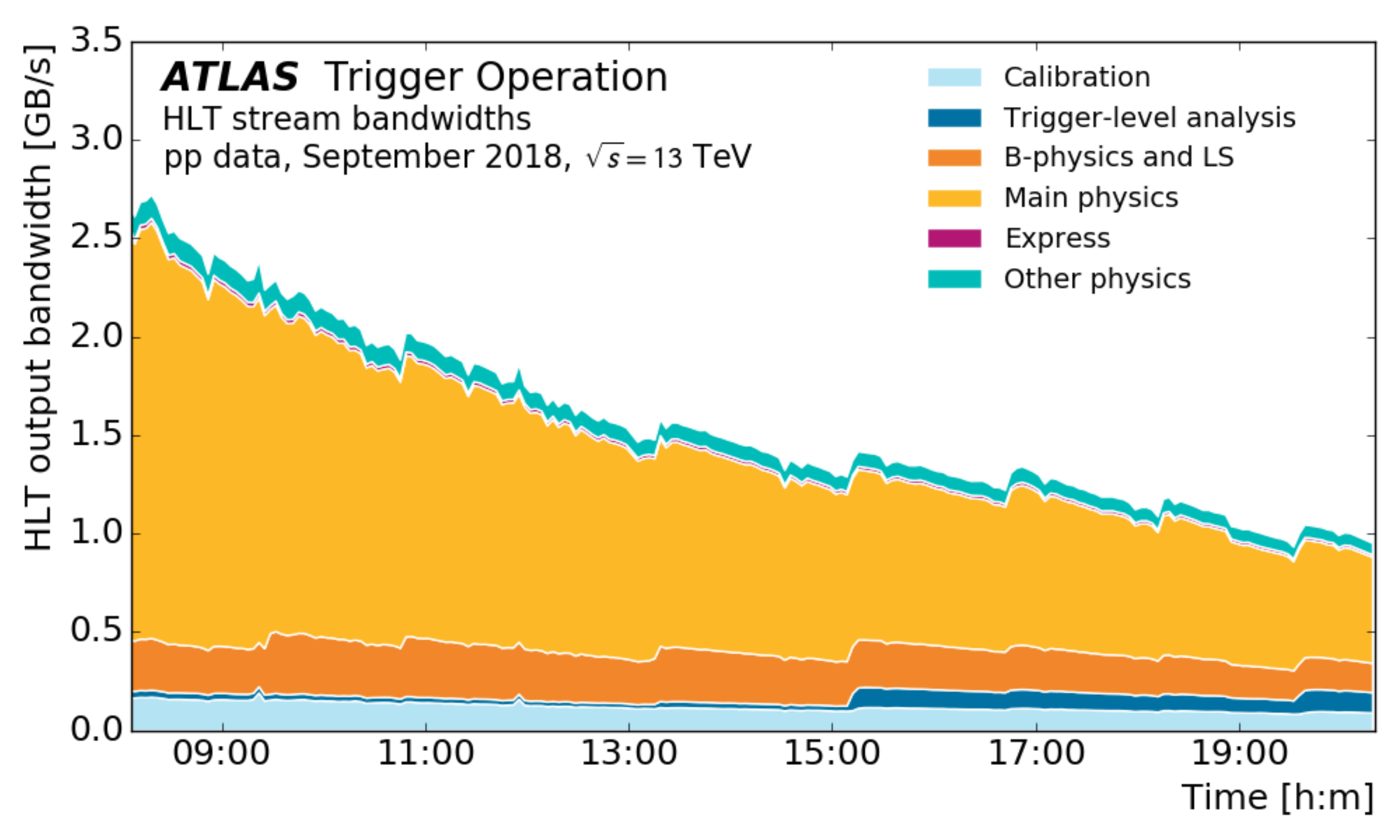}
\caption{Trigger stream rates (top) and output bandwidth at the HLT (bottom) as a function of time in a fill taken in September 2018 with a peak luminosity of $2.0\times\lumi{e34}$ and a peak pile-up of 56.
Presented are the main physics stream rate, containing all trigger chains for physics analyses; the BLS stream, containing trigger chains specific to $B$-physics analyses; the express stream, which records events at a low rate for data quality monitoring; other physics streams at low rate, such as beam-induced background events; the trigger-level analysis stream; and the detector calibration streams.
The monitoring stream is not reflected in the output bandwidth as the monitoring data are not written out to disk. The increase of the TLA HLT output rate is part of the end-of-fill strategy of the ATLAS trigger. At the end of the LHC fill, L1 and CPU resources are available to reconstruct and record additional events using lower-threshold TLA triggers.  During Run 2 the TLA stream was seeded by jet triggers and only the HLT jet information was saved. This increased the total HLT output rate, but did not significantly increase the total output bandwidth due to the small size of TLA events.}
\label{fg:HLTRatesBandwidth}
\end{center}
\end{figure}

Figure~\ref{fg:HLTRatesBandwidth} shows the HLT rates and output bandwidth as a function of time in a given run.
The apparent mismatch between rate and output bandwidth in some streams is due to the use of PEB techniques.
The increase of the TLA HLT output is part of the end-of-fill strategy of the ATLAS trigger.
Towards the end of the LHC fill, when the luminosity and the pile-up are reduced compared to their peak values, L1 bandwidth and CPU resources are available to record and reconstruct additional events using lower-threshold TLA trigger chains.
 
Table~\ref{tb:eventsizes} shows the average event sizes for the described streams. The event size of physics, express and BLS streams is comparable whereas the TLA stream event size is significantly smaller. The calibration stream size varies considerably depending on the purpose and what sub-detector information is written out.
 
\begin{table}
\centering
\caption{The average event sizes for the physics and express stream, the trigger-level analysis, calibration and $B$-physics and Light States (BLS) are presented in this table.}
\begin{tabular}{|l|c|}\hline
Stream & Average event size \\ \hline
Physics, express & 1 MB\\
Trigger-level analysis & 6.5 kB\\
Calibration & 1.3 kB to 1 MB \\
$B$-physics and light states &  1 MB\\
\hline
\end{tabular}
\label{tb:eventsizes}
\end{table}

\pagestyle{plain}
 
\section{\label{sec:special}Special data-taking configurations}
In addition to standard $pp$ and heavy-ion data-taking, the LHC programme includes a variety of short periods when the machine is operated with particular beam parameters, referred to as special data-taking configurations.
The special data-taking configurations provide data for detector and accelerator calibration as well as additional physics measurements in the experiments.
The specific LHC bunch configurations and related conditions (e.g.~lower number of paired bunches, change in the average number of pile-up interactions), detector settings (e.g.~subsystem read-out settings optimised for collecting calibration data) and desired trigger configuration have to be taken into account when preparing the trigger menu.
The preparation of these configurations can be quite extensive as they require a specific trigger menu which needs to be prepared and adjusted to comply with the imposed, usually tightened limits in rate, bandwidth and CPU consumption. 
In the following, three examples of special data-taking configurations, and the challenges that come with them, are discussed: runs with a low number of bunches, luminosity calibration runs and a configuration used to record enhanced minimum-bias data for future estimates of trigger rates and CPU consumption. 
 
\subsection{\label{sec:lownumberbunches}Runs with few bunches}
\begin{figure}
\begin{center}
\includegraphics[width=0.7\textwidth]{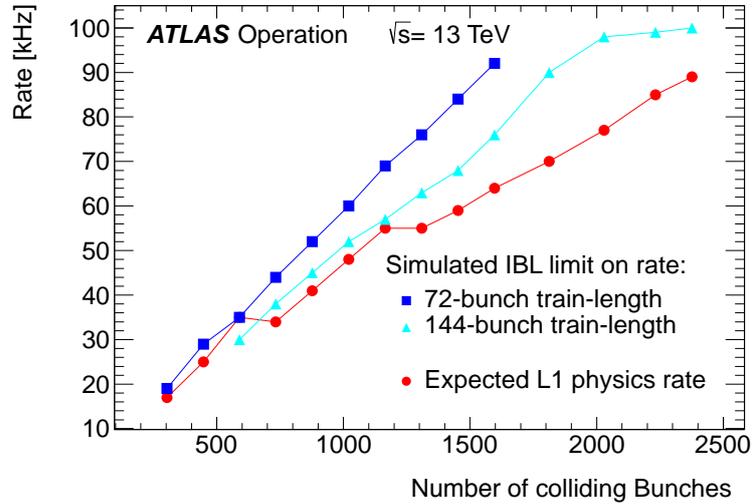}
\caption{The fixed-frequency veto limit to protect the innermost pixel detector of ATLAS (IBL) against irreparable damage due to resonant vibrational modes of the wire bonds has a direct impact on the maximum allowable rate of the first trigger level (L1).
This limit depends on the number of colliding bunches in ATLAS and on the filling scheme of the LHC beams.
This plot presents the simulated rate limits of the L1 trigger as imposed for IBL protection for two different filling schemes (in blue), and the expected L1 rate (in red) from rate predictions.
The steps in the latter indicate a change in the prescale strategy.
The rate limitation is only critical for the lower-luminosity phase, where the required physics L1 rate is higher than the limit imposed by the IBL veto.
The rate can be reduced by applying tighter prescales.}
\label{fig:L1limitIBL}
\end{center}
\end{figure}

Runs with a low number of bunches (e.g.~3, 12, 70, 300 bunches) usually occur during the periods of intensity ramp-up of the LHC after long or end-of-year shutdowns~\cite{Wollmann:2016ezf}.
While in most of these runs it is still desired to collect data for detector calibrations or for physics analyses that prefer low-luminosity conditions, they provide an operational challenge due to certain limits of the ATLAS detector which the trigger needs to take into account.
The most stringent limitation when a small number of bunches are grouped into small sets of bunches (\textit {bunch trains}) arises from events being accepted at L1 and the data being read out at the mechanical resonance frequencies of the wire bonds of the insertable B-layer (IBL) or the semiconductor tracker (SCT).
This can cause physical damage to the wire bonds.
The resonant vibrations are a direct consequence of the oscillating Lorentz forces induced by the magnetic field and cause wire bonds to break due to fatigue stress.
The resonant modes of the wire bonds lie at frequencies between 9 and 25~kHz for the IBL, which is of concern given the 11\,245~Hz LHC bunch revolution frequency.
The resonant modes of the SCT are less of a concern as they are typically above the maximum L1 trigger rate limit imposed by the IBL.
To protect the detector, a so-called fixed-frequency veto is implemented, which prevents read out of the detector upon sensing a pattern of trigger rates falling within a dangerous frequency range~\cite{AlvarezFeito:2010249, SCTwirebonds}.
The IBL veto provides the most stringent limit on the L1 rate in this particular LHC configuration.
To prepare trigger menus which respect this limit, the maximum affordable trigger rate is first determined by simulating the effect of the IBL veto.
If the expected rate from the nominal trigger menu is higher than the allowed rate, the menu is adjusted to reduce the rate to fit within the limitations.
Figure~\ref{fig:L1limitIBL} shows the simulated IBL rate limit for two different bunch configurations, together with the expected L1 trigger rate of the nominal physics trigger menu.
This rate limitation is only critical for the lower-luminosity phase, where the required physics L1 rate is higher than the limit imposed by the IBL veto.
In order to avoid impacting primary physics triggers, the required rate reduction is achieved by reducing the rate of the supporting trigger chains.
 
\subsection{Luminosity calibration runs}
Luminosity calibration runs are runs in which the absolute luminosity scale~\cite{ATLAS-CONF-2019-021} is being determined and the calibration of the different luminosity detectors is measured.
A precise measurement of the integrated luminosity is a key component of the ATLAS physics programme, in particular for cross-section measurements where it is often one of the leading sources of uncertainty.
The luminosity measurement is based on an absolute calibration of the primary luminosity-sensitive detectors in low-luminosity runs with specially tailored LHC conditions using the van der Meer (vdM) method~\cite{vanderMeer:296752}.
The luminosity calibration relies on multiple independent luminosity detectors and algorithms, which have complementary capabilities and different systematic uncertainties. 
One of these algorithms is the counting of tracks from the charged particles reconstructed in the inner detector in randomly selected bunch crossings.
Since the different LHC bunches do not have the exact same proton density, it is beneficial to sample a few bunches at the maximum possible rate.
For this purpose, a minimum-bias trigger~\cite{ATLAS-CONF-2010-068} selects events for specific LHC bunches and uses partial event building to read out only the inner-detector data. The data are read out at about 5~kHz for five different LHC bunches defined in the specific bunch group of the bunch group set used in the run.

\subsection{\label{sec:ratePred}Enhanced bias runs}
Certain applications such as HLT algorithm development, rate predictions and validation (described in Section~\ref{sec:ReleaseValidation}) require a dataset that is minimally biased by the triggers used to select it.
The {\textit {Enhanced Bias}} (EB) mechanism allows these applications to be performed utilising dedicated ATLAS datasets.
These datasets contain events only biased by the L1 decision, by selecting a higher fraction of high-$p_\mathrm{T}$ triggers and other interesting physics objects than would be selected in a \textit{zero bias} sample (i.e.~a sample collected by triggering on random filled bunches).
To collect the EB dataset, a specific trigger menu is used which consists of a selection of representative L1 trigger items spanning a range from high-$p_\mathrm{T}$ primary trigger items to low-$p_\mathrm{T}$ L1 trigger items, plus a random trigger item to add a zero-bias component for very high cross-section processes. The random trigger item corresponds to a random read-out from the detector on filled bunches and therefore corresponds to a totally inclusive selection.
The bias from the choice of items in the EB trigger menu is invertible, which means that a single weight is calculable per event to correct for the prescales applied during the EB data-taking.
This weight restores an effective zero-bias spectrum.
The recorded events are only biased by the L1 system, no HLT selection is applied beyond the application of HLT prescales to control the output rates.
The EB trigger menu can be enabled on top of the regular physics menu, adding a rate of 300~Hz for the period of approximately one hour in order to record around one million events.
This sample contains sufficient events to accurately determine the rate of all primary, supporting and backup trigger chains which together make up a physics trigger menu.

\pagestyle{plain}
 
\section{Condition updates in the HLT}
\label{sec:conditions}
 
The HLT event selection is driven by dedicated reconstruction and selection algorithms.
The behaviour and performance of some of those algorithms depend on condition parameters, or \textit{conditions}, which provide settings, such as calibration and alignment constants, to the algorithms.
Conditions are valid from the time of their deployment, and until their next update.
Depending on the nature of the conditions, these updates can be frequent.
While most conditions are updated only between runs and often much less frequently, some are volatile enough to require updates during ongoing data-taking.
In the ATLAS experiment, all conditions data and their interval of validity are stored in the dedicated COOL database~\cite{COOL}.
This section describes those special conditions and the procedure that was introduced to configure them consistently and reproducibly across the HLT farm.
 
\subsection{Conditions updates within a run}
 
\subsubsection{Online beam spot}
 
Many criteria employed in the event selection are sensitive to the changes in the transverse and (to a lesser extent) longitudinal position and width of the LHC beams, also referred to as the \textit{beam spot}~\cite{TRIG-2011-02}.
The parameters of the beam spot are important inputs for the selection of events with $B$-hadrons, which have a long lifetime, 
typically decaying a few millimetres from the primary proton--proton interaction vertex.
Since the beam-spot parameters are not constant within a run,
they are continuously monitored and updated during data-taking if there are large enough deviations from the currently used values.
The beam spot is estimated online by collecting the primary-vertex information provided in histograms created by HLT algorithms executed on events selected by L1 jet triggers.
These histograms are then collected by an application external to the HLT, and the beam-spot position and tilt are determined in a fit.
For every new LB, the beam-spot application reads in the histograms of the last few LBs; usually at least four or five LBs are required in order to acquire enough statistics to perform an initial beam-spot fit.
If the fit is successful, the conditions update procedure for the beam spot is started if any of the following is true:
a) the beam position along any axis relative to the beam width has changed by more than 10\% and the significance of this change is larger than two, b) the width has changed by more than 10\% and the significance of this change is larger than two, or c) the precision of either the beam position or the width has improved by more than 50\%.

\subsubsection{Online luminosity}
 
Since the bunches in the LHC arrive in trains, there are several consecutive bunch crossings with collisions followed by a gap between the trains with empty bunch crossings.
Additionally, the bunches in the train have slightly different bunch charges, which means that the luminosity for each bunch can be different from the average luminosity across the full train.
The signals from energy depositions in the liquid-argon calorimeter span many bunch crossings, affecting the energy reconstruction of subsequent collision
events.
Therefore, the signal pedestal correction that is applied during the energy reconstruction depends on the per-bunch luminosity of the event bunch itself and that of the surrounding bunches~\cite{LARG-2013-01}.
The LUCID detector continuously monitors the overall and per-bunch luminosity, while a separate application compares it with the currently used luminosity values and starts the conditions update procedure for the luminosity if the average luminosity deviates by more than 5\%.
The per-bunch pile-up values are also used in pile-up-sensitive algorithms to correct for the bunch-crossing dependence of the calorimeter pulse-shapes~\cite{ATL-DAQ-PUB-2017-001}, and for reconstruction of electron~\cite{TRIG-2018-05} and hadronic $\tau$-lepton decay~\cite{ATLAS-CONF-2017-061} candidates.
 
\subsubsection{Updates of trigger prescales}
\label{sec:prescales}
Prescales can be used to either adjust or completely disable the rate of an item/chain, or to only allow its execution after the event has already been accepted (so-called \textit{rerun} condition).
Being able to change trigger prescales during an ongoing run is extremely valuable in providing luminosity-dependent trigger configurations during data-taking and therefore in making optimal use of the available resources.
As opposed to L1, where prescales are applied centrally by the CTP, the update of HLT prescales needs to be synchronised across the entire HLT farm.
In addition, the HLT nodes need to be able to access the history of prescale sets used for the currently ongoing run to be able to apply the correct prescales due to the processing delay in the HLT.
For this purpose, a table holding the mapping between LBs and prescale keys is recorded in the trigger database (see Section~\ref{sec:TriggerConfig}), and propagated to the HLT farm using the same mechanism as the other conditions updates.
Any change of prescale key is also recorded in COOL for use in offline analysis.

\subsection{Conditions update procedure}
 
\begin{figure}
\centering
\includegraphics[width=0.48\textwidth]{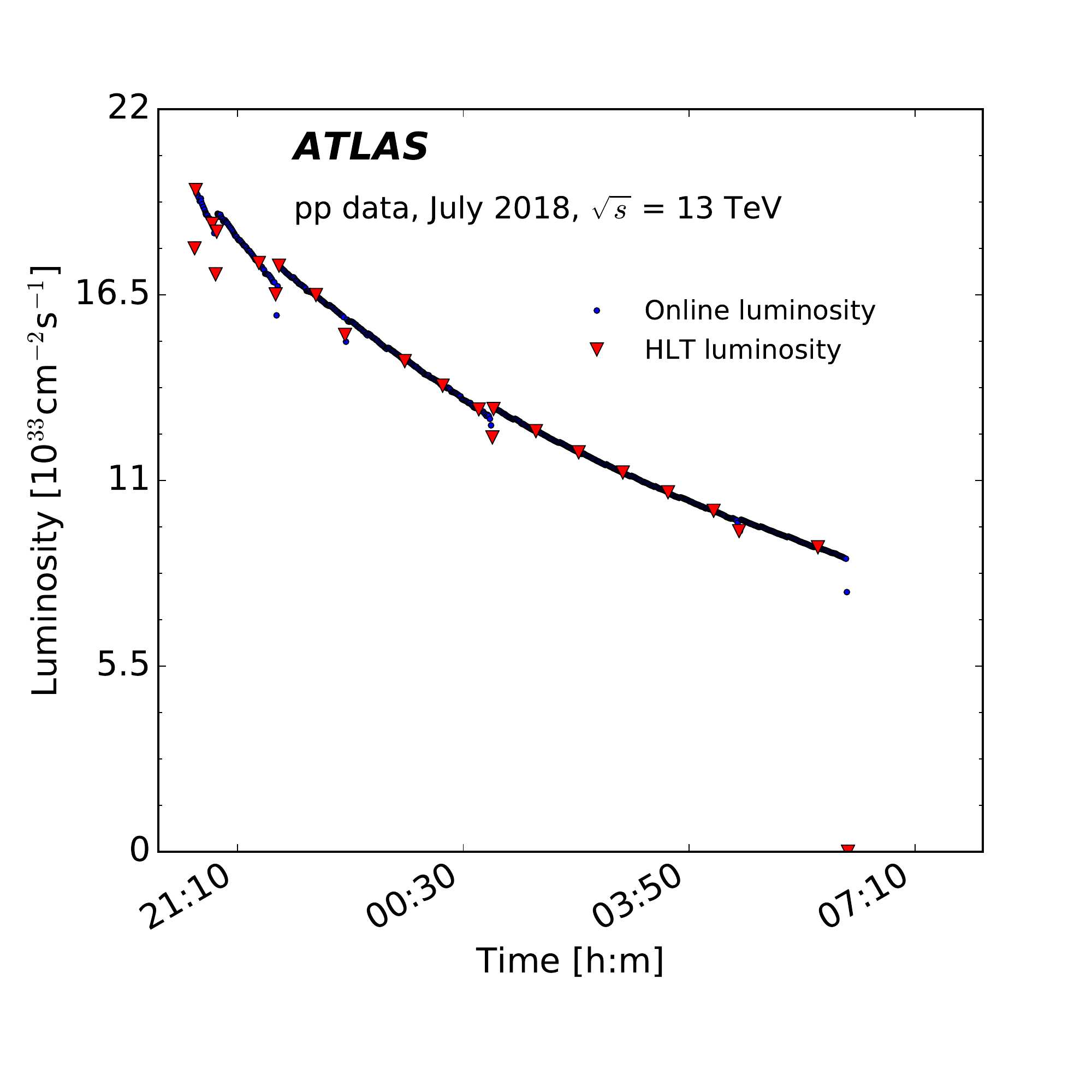}
\includegraphics[width=0.48\textwidth]{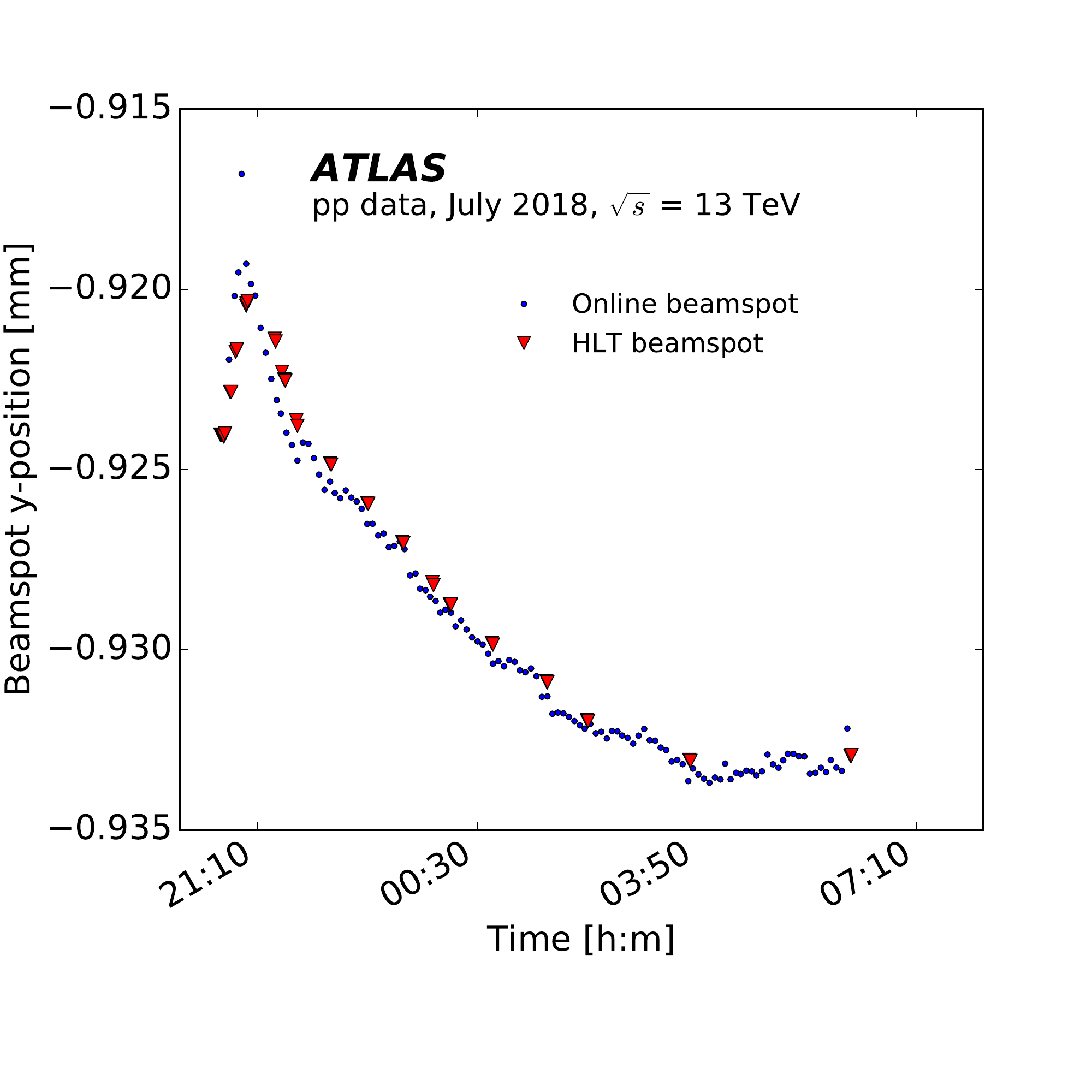}
\caption{The live values (blue circles) and values used in the HLT (red triangles) of the luminosity (left) and the $y$-position of the beam spot (right) for a typical $pp$ run. The sudden increases in the luminosity are from the LHC re-optimisation of the colliding beams.}
\label{fig:lumiBeamSpot}
\end{figure}

To update conditions during the run the new data are written to COOL with a validity range starting in the next LB and the HLT processes are informed through markers in the event data itself that the new information has to be fetched from COOL.
Distributing the update markers through the data path rather than the control path ensures that all HLT nodes receive it.
The HLT algorithms that use such volatile conditions data, i.e.~online beam-spot or luminosity information, receive a handle to the relevant conditions object and the framework takes care of providing the correct conditions data for every event (see Ref.~\cite{Winklmeier:1271370} for more details).
An example showing the luminosity and the $y$-position of the beam spot used in the HLT compared with their respective live values  is shown in Figure~\ref{fig:lumiBeamSpot}.
 
\pagestyle{plain}
 
\section{Trigger configuration}
\label{sec:TriggerConfig}
 
The settings of both the hardware and software components of the ATLAS trigger system must be accurately recorded to ensure the correct interpretation of the trigger in offline analysis and the reproducibility of the trigger behaviour in the recorded data.
The ATLAS trigger configuration system uses a relational database, the TriggerDB, to store these configuration data in a stable and accessible manner. L1Muon and L1Calo maintain additional configuration and condition databases, while the CTP stores most of its information in the TriggerDB.
The Run-2 ATLAS trigger configuration system is described in detail in Ref.~\cite{Martyniuk:2016459} and is summarised in the following.
 
The upload of the configuration to the TriggerDB is handled by the TriggerTool via its graphical user interface (GUI). The TriggerTool GUI also makes it possible to browse the TriggerDB configurations and to edit existing configurations, which results in new configurations that maintain reproducibility.
 
The definition of the L1 configuration contains: the L1Topo trigger menu, a list of algorithms and their parameters used in the L1Topo system; the muon and calorimeter $p_\mathrm{T}$ thresholds and isolation requirements for each L1 trigger object; the L1 trigger menu, a list of L1 trigger thresholds and objects which are combined into L1 trigger items; the CTP firmware logic, which is derived from the L1 trigger menu; and global settings, such as the L1Calo isolation settings.
To accurately describe the HLT software configuration, two components are needed: the HLT trigger menu and parameters needed to configure the HLT algorithms and other software components employed by the trigger chains. The software release version that was used to produce this configuration is also stored in order to provide a reference to the algorithm implementation that is applicable for the intended release (the original or a later release version).
 
In addition to the configuration, which is fixed during a run, the TriggerDB also stores information that can be updated during physics data-taking (see Section~\ref{sec:conditions}), the prescales to be used with each L1 item or HLT trigger chain and the bunch group set (see Section~\ref{sec:BunchGroups}).
 
\subsection{The trigger database design}
\begin{figure}[tb]
\centering
\includegraphics[width=0.9\textwidth]{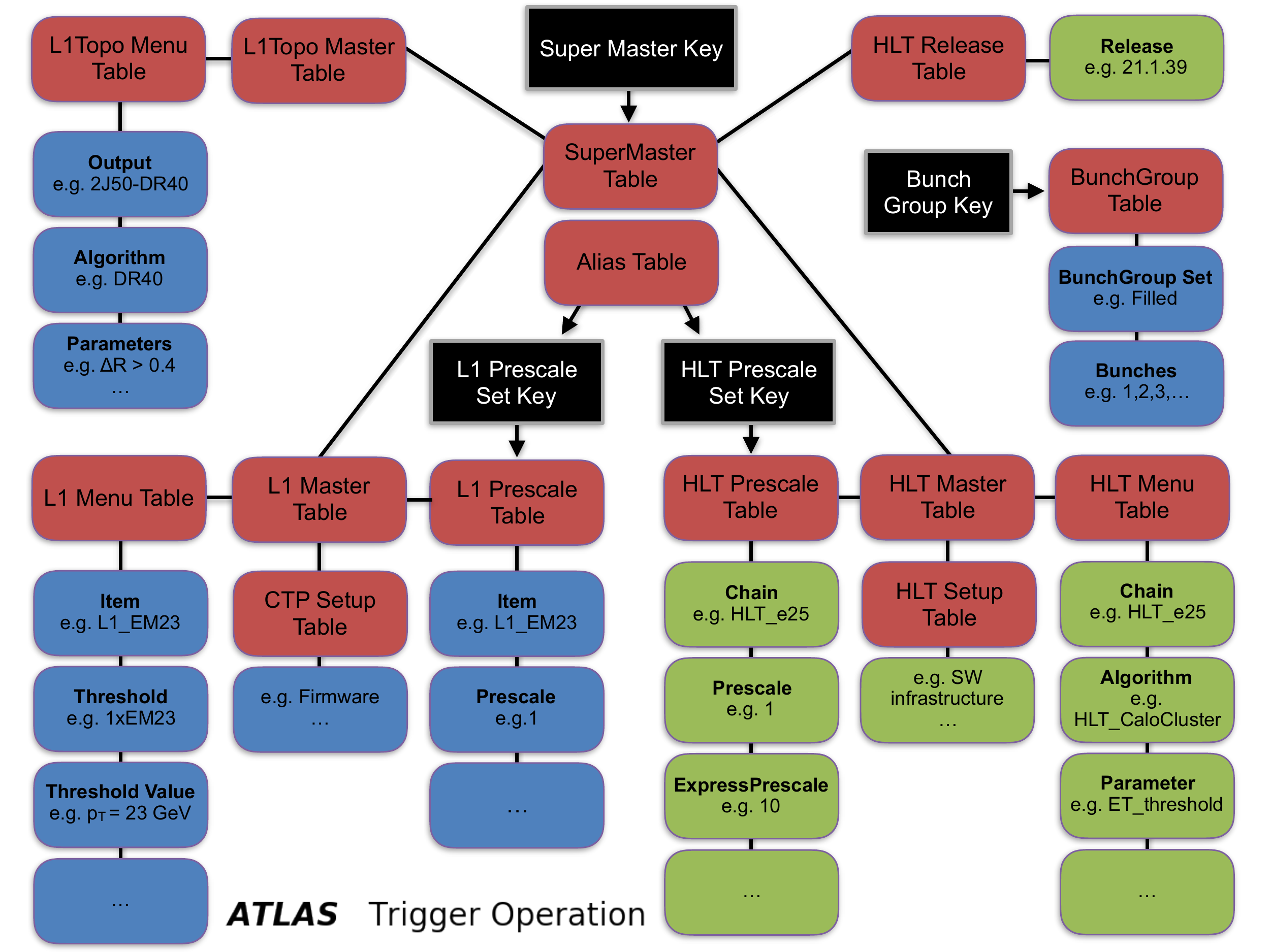}
\caption{A simplified representation of the database structure for the trigger configuration highlighting the primary keys and the relationship between a selection of tables. The primary keys (black rectangles) are associated with an index of a table (rectangles). Most of the top-level table structure is shown (red rectangles) with each of these linking to further tables (for example a trigger chain). These linked tables contain various objects (for example a chain name) and subsequent links to still further tables (for example the algorithm for the given trigger chain) as demonstrated by the blue (L1) or green (HLT) rectangles.}
\label{fig:DB}
\end{figure}

The TriggerDB~\cite{Chavez:2015kxy} is an Oracle relational database schema in which all tables have an integer primary key corresponding to a unique combination of a name and version for the object and any dependent object that is stored.
The primary key of the top-level table is known as the Super Master Key (SMK) and uniquely defines the L1 and HLT configuration.
The SMK together with the Bunch Group Key (BGK), associated with the bunch group set, and Prescale Set Keys (PSK) for L1 and the HLT uniquely define the trigger configuration used to record a run.
These keys are saved in COOL for each LB for use by offline analyses.
A simplified representation of the database structure for the trigger configuration is shown in Figure~\ref{fig:DB}.
 
There are several instances of the TriggerDB hosted on the CERN Oracle servers, each for different use cases.
The online TriggerDB is accessed by custom C++ configuration classes to load the configuration into the L1 and HLT systems at the start of a run.
The online DB is only accessible from within the ATLAS control room network, but there is a read access clone replicated externally.
This is needed to be able to run the trigger simulation using the same configuration classes as online.
Other TriggerDB instances are used to produce MC simulation and for trigger reprocessing tasks.
 
The design of the TriggerDB is based on two principles: to avoid duplication of records in the database and to maintain every configuration ever used to record data.
Link tables are used to allow the same primitive to appear in many configurations and similar primitives to appear within the same configuration. This removes the need to replicate the primitives themselves.
If there is a change to any part of the configuration then it will always result in a new SMK, with only the updated information being saved.
The TriggerDB is also used by a web page where the configuration can be browsed for a specific set of trigger keys, or looked up for a specific run.

\subsection{The TriggerTool}
The commonly used way to upload the trigger configuration into the database is through the TriggerTool GUI~\cite{Bell:2009jca}, a Java project developed in the NetBeans~\cite{netbeans} integrated development environment.
Upon launching the tool, the user has to choose their access role 
and the database connection. 
 
\begin{figure}[tb]
\centering
\includegraphics[width=0.9\textwidth]{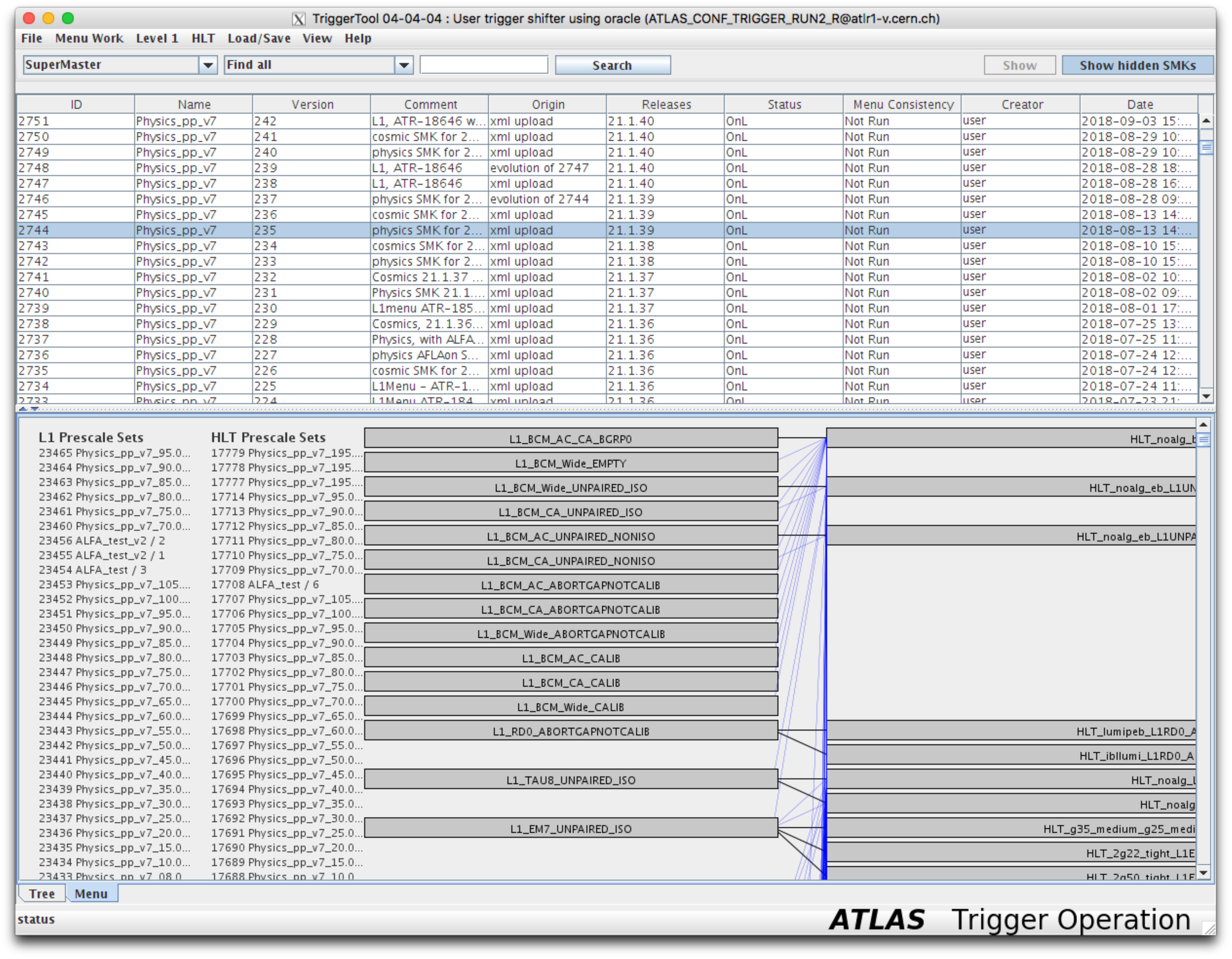}
\caption{The main panel of the TriggerTool, displaying all available Super Master Keys (SMKs) which are currently contained in the connected TriggerDB with a minimal set of information such as its creator, time of creation and software release with which it was created. In the bottom left part, some of the available prescales keys are shown. In the bottom right, a visual display of the trigger chains and their corresponding L1 items is displayed.}
\label{fig:triggertool}
\end{figure}

\begin{figure}[tb]
\centering
\includegraphics[width=0.9\textwidth]{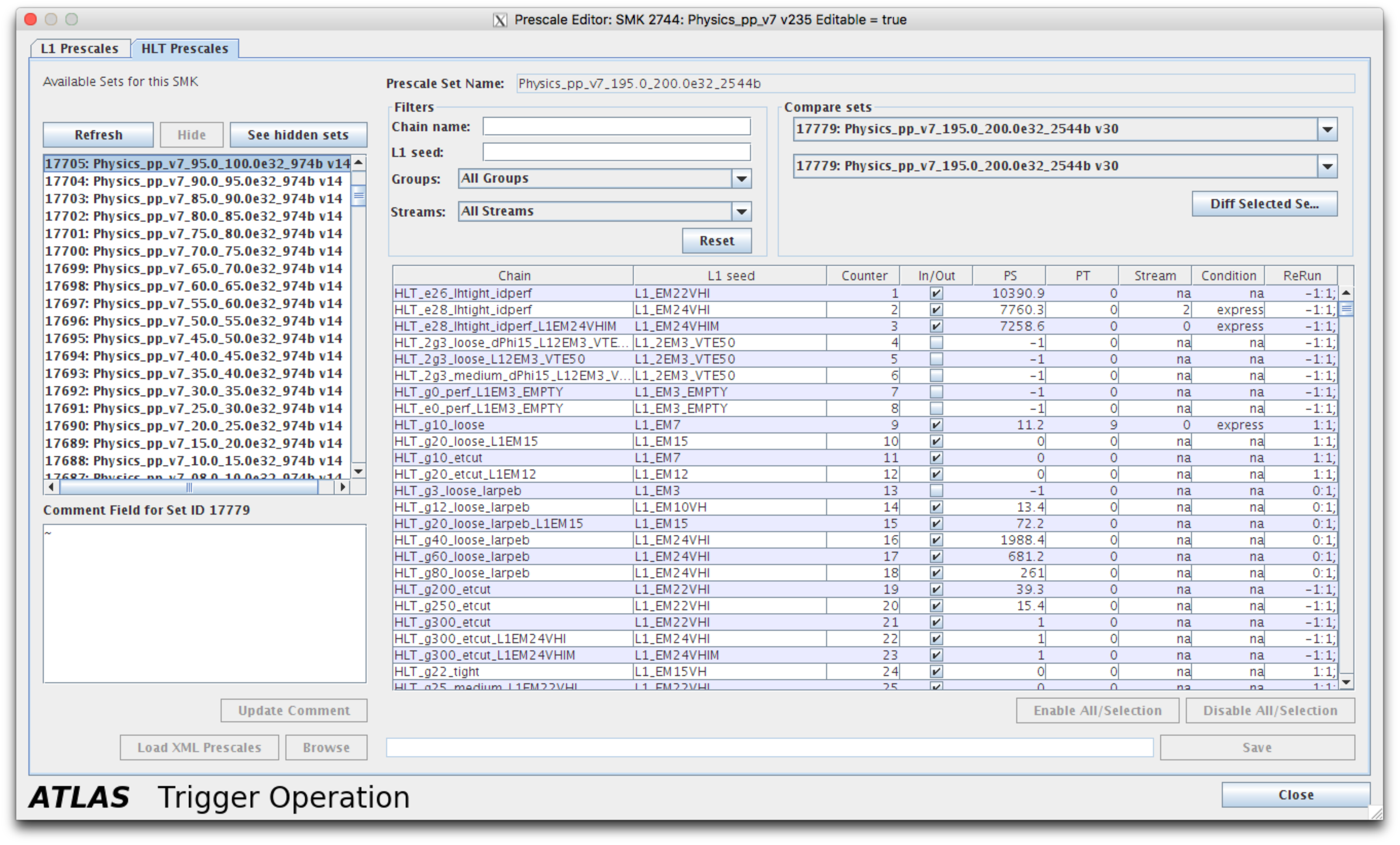}
\caption{The HLT prescale tab of the TriggerTool GUI with the list of available Prescale Set Keys for use with the selected Super Master Key.
When a prescale set is chosen from the list, the related settings for each chain are shown in the table. Similar information for L1 items is available under the L1 prescales tab.}
\label{fig:hltprescales}
\end{figure}

Within the TriggerTool, a variety of options are available to display and edit the trigger configuration through a range of panels.
The main panel is shown in Figure~\ref{fig:triggertool}, which shows the configurations which are currently contained in the connected TriggerDB, as well as the L1 and HLT PSKs that are associated with a given SMK.
Selecting one of these prescale keys loads the details of that particular key, displaying the L1 items or HLT chains, their prescales and whether they are enabled or not. An example display for the HLT prescales can be seen in Figure~\ref{fig:hltprescales}.
The table can be used to edit the prescale or enable/disable an item. Changes are then saved as a new prescale key and associated with the same SMK.
Various options for displaying differences between prescale sets of the same SMK are given. 
There are additional panels to view or edit the configuration, which is especially useful for understanding algorithm settings or stream configurations. 
 
In preparing for and during data-taking, the TriggerTool is routinely used by the trigger menu and trigger online on-call experts to prepare the trigger configuration for the upcoming runs or to generate new trigger configurations during an ongoing run. Any change in prescales or bunch group sets can be deployed during a run, while a change in the trigger algorithm parameters (which leads to the generation of a new SMK) requires stopping the ongoing run and starting a new run.

\subsection{The TriggerPanel}
 
\begin{figure}[tb]
\centering
\includegraphics[width=0.9\textwidth]{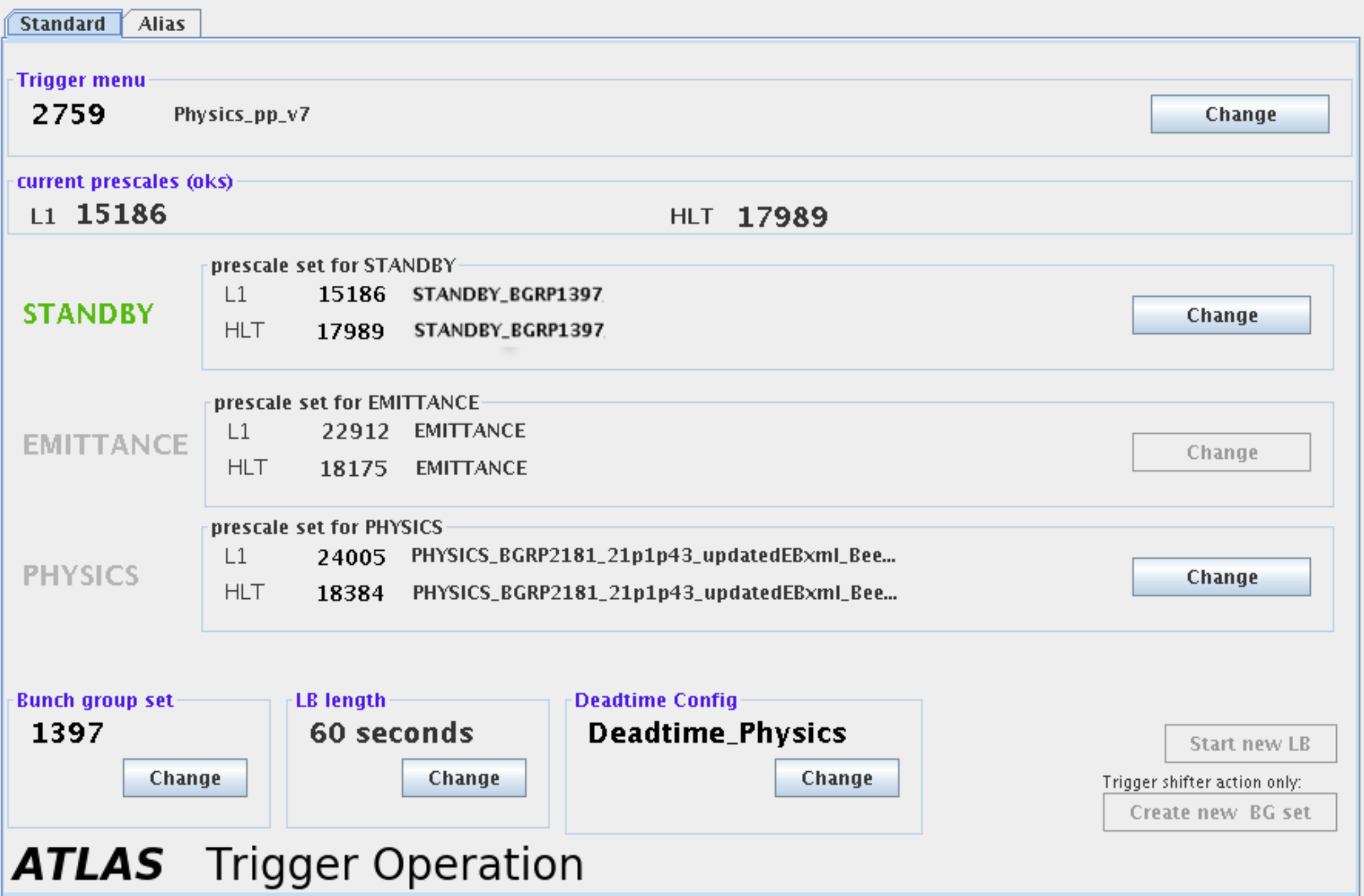}\\
\vspace{0.5cm}
\includegraphics[width=0.9\textwidth]{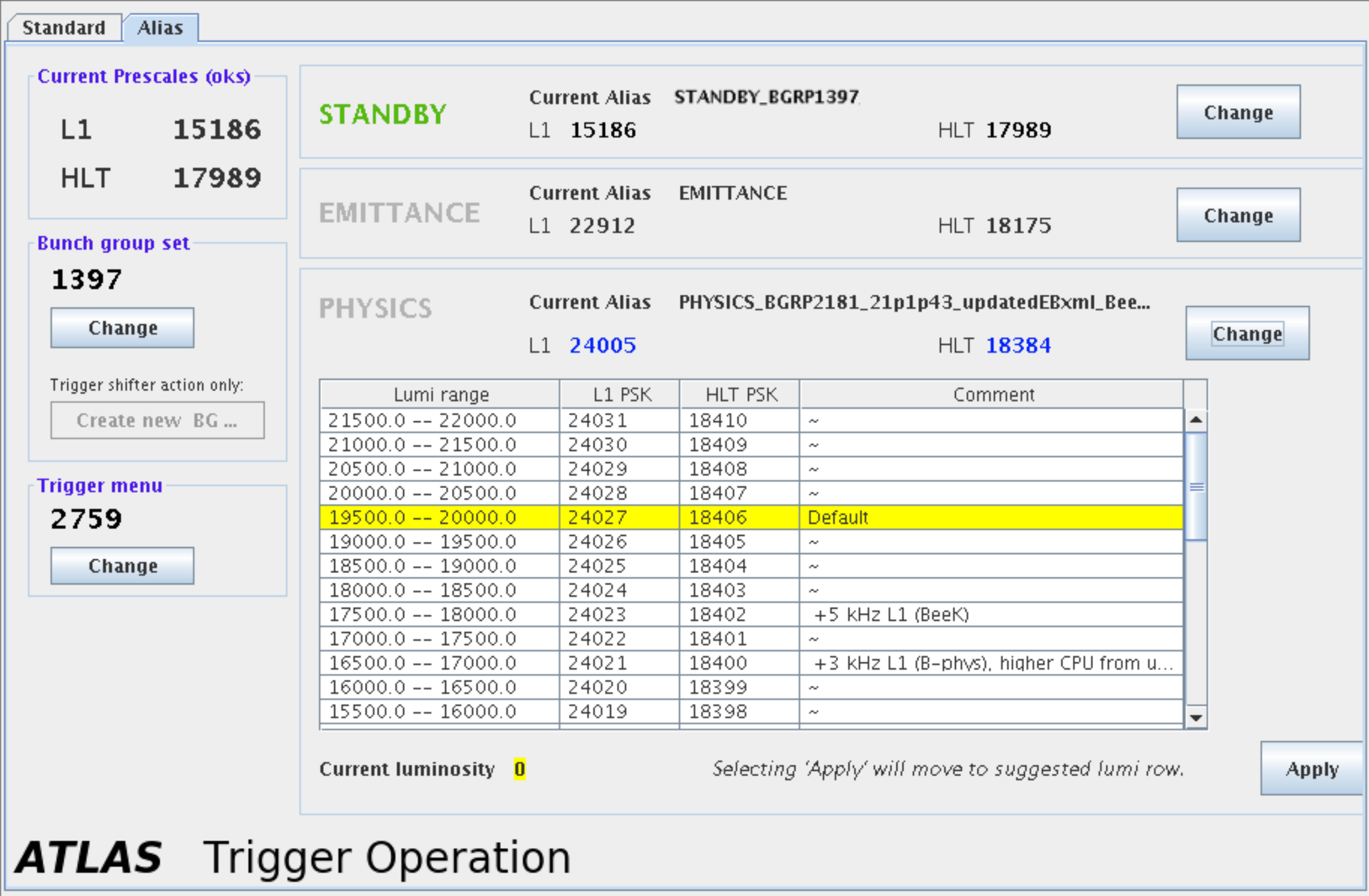}
\caption{The Standard (top) and Alias (bottom) tabs in the Trigger Panel. The following settings are displayed: the current prescales, the bunch group set and the Super Master Key (under Trigger menu), as well as the Prescale Key Sets (PSKs) to be used for various data-taking cases (Standby, Emittance, Physics). In the alias panel, for the case of physics data-taking, the alias table is displayed with the relevant luminosity range, L1 and HLT PSK and additional comment field as a suggestion of which prescale keys to use. }
\label{fig:panel}
\end{figure}

The change of the SMK and PSKs is handled by the TriggerPanel, which is included as part of the TDAQ software controlling the finite-state machine of ATLAS in the integrated run control GUI~\cite{Avolio_2011}.
The TriggerPanel displays all prescale keys that are associated with the loaded SMK, along with which keys are currently in use, as can be seen in Figure~\ref{fig:panel}.
The prescale keys can be set for different data-taking scenarios: standby, emittance scans, or physics.
To facilitate the selection of a corresponding L1 and HLT PSK assigned to a particular luminosity range, they can be linked via alias tables in the TriggerDB for this luminosity range.
Once an alias is selected in the TriggerPanel, it allows the PSKs to be suggested automatically, based on the current measured luminosity, and these are used only when confirmed by the shifter.
When the prescale keys are changed in the TriggerPanel, the keys are sent to the CTP, to then be forwarded to the HLT as described in Section~\ref{sec:prescales}.
 
\subsection{Automatic prescaling of L1 trigger items}
The Auto-Prescaler~\cite{GormThesis} is an application that was developed to automatically create and apply new prescale keys to L1 trigger items which are particularly sensitive to noise from sub-detectors. These can cause high rates, potentially exceeding the limits of the trigger system.
The Auto-Prescaler consists of two parts, a rule checker responsible for evaluating the automation rules against the available monitoring data and requesting any prescale changes, and an orchestrator which is responsible for processing the prescale changes, generating a new L1 prescale set if needed and automatically applying the new prescale set.
The prescales are calculated in multiples of an integer value, which is configurable for each L1 trigger item.
The functionality of the Auto-Prescaler was extended during \runii to also reduce or remove prescales applied to L1 items during non-physics beam modes.
This is used for recording beam background data during periods when no collisions are expected. 
The prescales set by the Auto-Prescaler automatically revert to the previous setting once the beam mode has changed or the rate spikes from detector noise have subsided.
 
\clearpage
\pagestyle{plain}
 
\section{Online release validation}
\label{sec:ReleaseValidation}
The online trigger software release cycle includes development, validation and deployment of the software to ensure the reliability and predictability of the performance of the HLT. This release is kept separate from the offline reconstruction release to be able to make use of new developments and fixes for data-taking.
The final validation of the trigger release is done using a suitable EB dataset (see Section~\ref{sec:ratePred}) and mimicking the configuration and execution during the data-taking.
The performance evaluation covers aspects of low-level memory and CPU requirements as well as distributions and efficiencies of high-level physics quantities which are analysed by a range of experts.
This multifaceted task is crucial for the successful and efficient operation of the trigger system.
More details of the development, validation and integration cycle can be found in Ref.~\cite{Keyes:2239296}.
Throughout \runii, about 80 software releases were deployed online for physics data-taking.
 
The trigger software validation cycle is typically carried out weekly for the most recent trigger development release as well as in preparation for changes in running conditions, e.g.~for heavy-ion runs or new pile-up conditions.
This cycle requires a coordinated effort between many groups and experts.
The process of software development is mainly driven by the improvement of algorithms, the addition of new trigger chains, as well as addressing occasional problems in the software that have been observed online or offline.
New software changes are validated daily by automatic nightly tests. These tests run on a few EB data events and are meant to catch software bugs. They are monitored by the online release on-call expert.
Typically, a larger-scale validation is carried out once a week, running the new software meant to be deployed online over about 1~million EB data events.
This reprocessing is used to identify rare bugs as well as to assess the performance of the new software in terms of trigger selection efficiencies and CPU usage.
The reprocessing is also used to estimate the rates of individual trigger chains, groups of chains and the entire menu after correcting for the prescales used during collection of the EB data~\cite{ATL-DAQ-PUB-2016-002, Martin:2237426}.
 
\begin{figure}
\centering
\includegraphics[width=0.49\textwidth]{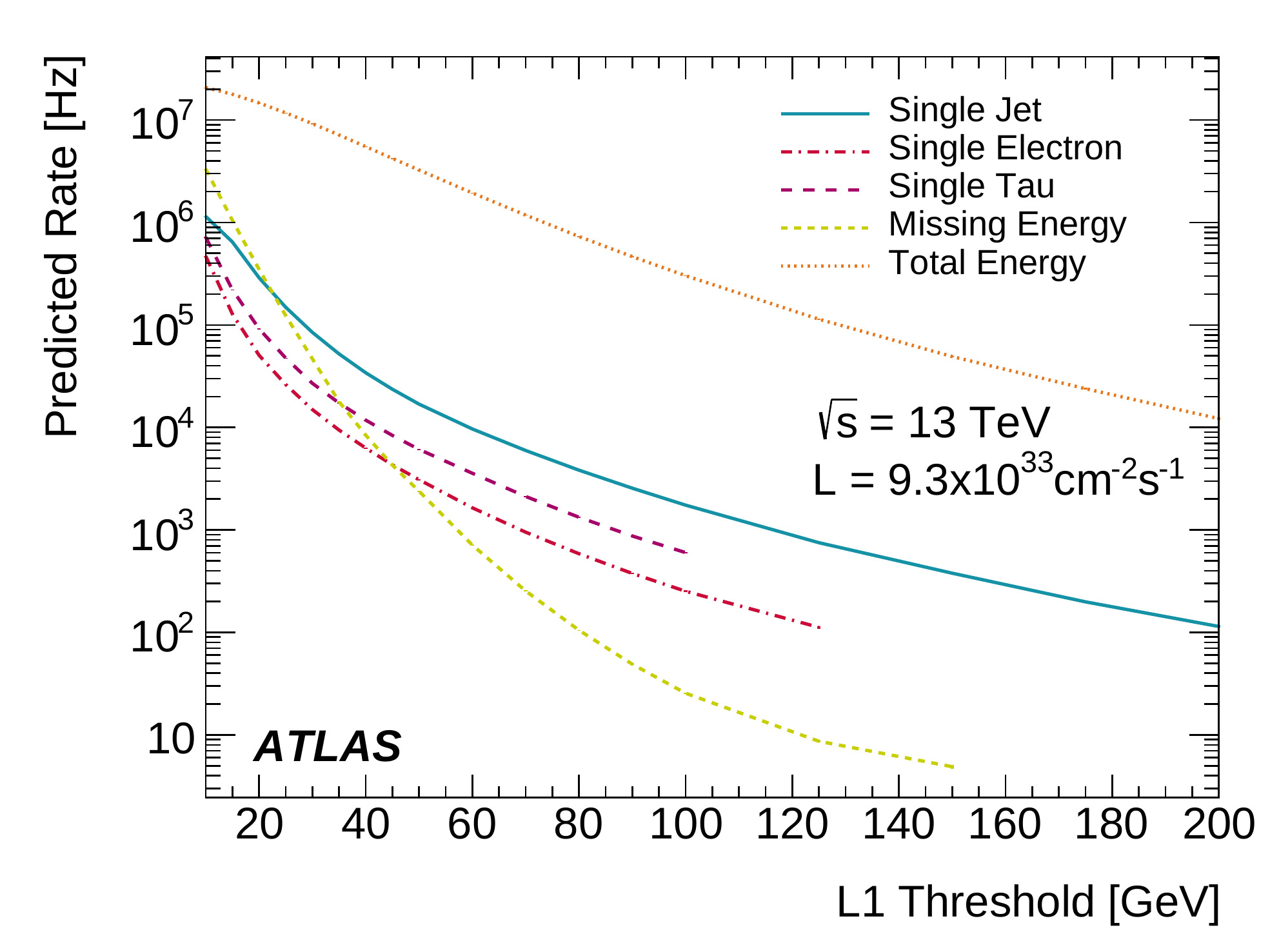}
\includegraphics[width=0.49\textwidth]{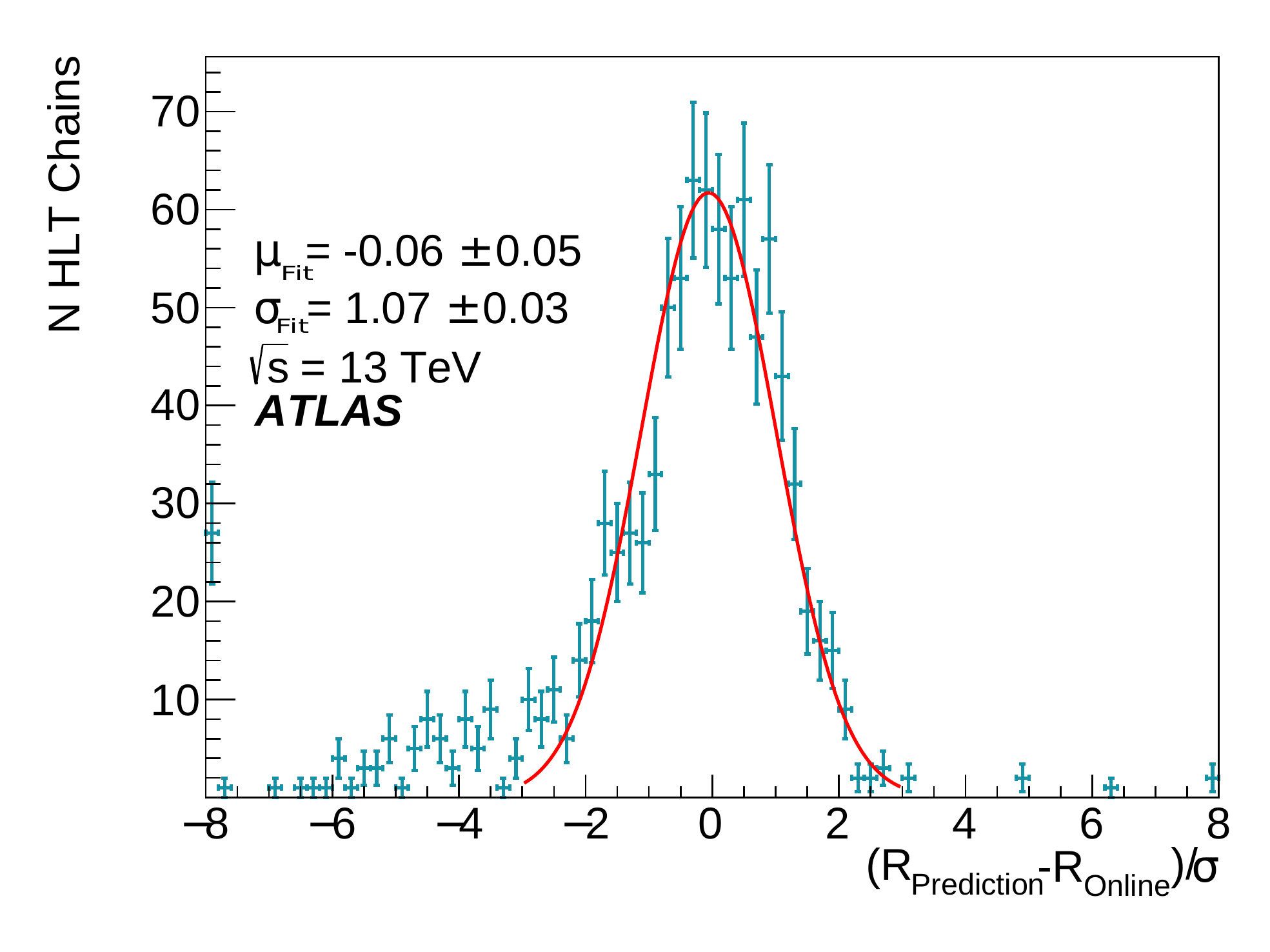}
\caption{Two examples of the EB rate predictions~\cite{Martin:2237426}. Left: Predicted rates as a function of $p_\mathrm{T}$ for various L1 triggers. Right: Pull distribution in predicted rates of 957 HLT chains from EB data normalised to the combined statistical error ($\sigma$), and fitted by a Gaussian function with mean, $\mu_{\mathrm{Fit}}$, and width, $\sigma_{\mathrm{Fit}}$, where $\mathrm{R_{Prediction}}$ and $\mathrm{R_{Online}}$ refer to the rates from prediction and data-taking for the same luminosity, respectively. Error bars include statistical uncertainties only.}
\label{fig:enhancedbiasperf}
\end{figure}

Two examples of the EB rate predictions are given in Figure~\ref{fig:enhancedbiasperf}. In Figure~\ref{fig:enhancedbiasperf} (left), predicted rates are shown versus transverse energies expressed as L1 thresholds. The smooth spectra, which are obtained via the weighting procedure, illustrate the statistical power of the data sample over several orders of magnitude in rate. In Figure~\ref{fig:enhancedbiasperf} (right), the HLT rate predictions are compared with online rates for 957 trigger chains of the trigger menu with a non-zero rate at the time of data-taking at the same luminosity. The recorded rates are corrected for prescales applied at L1 and/or the HLT, and the difference in rate between the prediction and the recording is normalised to the combined statistical uncertainty from both samples. The Gaussian fit in the range $-3$ to 3 indicates that the prediction for the majority of HLT chains is normally distributed. For a small number of chains where the predicted rate was too low, a tail to the negative significance is visible. This is due to a small bias arising from the chosen set of L1 seeds of the EB dataset in addition to luminosity scaling assumptions. Since the enhanced bias data are collected over a one hour period, the mean number of pile-up interactions changes throughout the sample and this introduces another small bias to chains whose rate is not linear with respect to the instantaneous luminosity, due to pile-up sensitivity. The mean fractional statistical uncertainty per HLT chain is 10\% for the predicted rates and 2\% for the recorded rates. For the total HLT rate, the statistical uncertainty of the prediction is 1\%.
 
The processing of the trigger software on the enhanced bias dataset is run on the Worldwide LHC Computing Grid~\cite{Worldwide:1997398} using the ATLAS production system, Prodsys2~\cite{Borodin:2015kfq}, and can be monitored and managed through a web interface.
Using the ATLAS Metadata Interface (AMI)~\cite{Albrand_2008}, the software validation expert manages and stores the configurations and parameters of the data reprocessing.
The reprocessing runs the HLT algorithms over raw data files, which is the same format as produced by the trigger system during data-taking, and produces another raw data file containing the new trigger decisions.
In addition to reprocessing with the HLT algorithms, it was also possible to run the L1 simulation on the raw data to obtain updated L1 decisions when new L1 trigger items were introduced in the trigger menu.
The raw data files containing the new trigger decisions are then processed with the ATLAS production reconstruction software to produce the output necessary for validation.
This process of rerunning the trigger and offline reconstruction on about 1M EB events takes around 24~hours and typically requires on the order of 7000~hours of CPU time, with each job using roughly 4~GB of virtual memory.
Throughout Run 2, the HLT reprocessing used about 0.2\% of the ATLAS grid resources.
 
The performance evaluation of a new release is carried out by signature-specific experts who compare various results with those from a previously validated release.
The offline and online monitoring is used to sign off a new release similar to the procedure used for the quality assessment of the collected data (see Section~\ref{sec:OfflineMonitoringDQ}).
Once the release is validated, it is deployed to the online machines for data-taking and further tested in an ATLAS test run during an LHC inter-fill period to check that there are no problems configuring the HLT with the new release in the online environment and that the release is compatible with software being used by other subsystems.

\pagestyle{plain}

\section{Debug stream processing}
\label{sec:DebugStream}
Sometimes the HLT is unable to make a decision on whether to accept an event due to processing failures.
Such failures can arise from algorithms crashing, time outs, missing data, etc. In such cases, the events are grouped by their failure type and recorded into a corresponding debug stream in order to be further studied offline, and in many cases recovered.
 
The failure types and corresponding debug streams are:
\begin{itemize}
\item \textbf{HLT Timeout}: Configurable limits are set on an event's processing time. To allow a graceful termination of the processing, a \textit{soft} timeout is applied which corresponds to 95\% of the value of the \textit{hard} timeout. If an event exceeds the soft timeout limit, the remaining algorithms are skipped, a partial HLT result\footnote{The HLT result contains a header summarising the overall trigger decision, plus more detailed intermediate information from the event processing, e.g.~amongst other information, the navigation data structure and some of the features which are useful for debugging.} is recorded, and the event is sent to the HLTTimeout stream. These events are recoverable offline where there is no limit on the processing time except in the cases where timeouts are caused by infinite loops which will require further fixes.
\item \textbf{Processing Unit Timeout}: If events from the soft timeout limit still do not finish processing within the hard timeout limit, they are sent to the PUTimeout stream. The settings for the hard timeout were 3 minutes in 2015, 5 minutes in 2016 and 7 minutes in 2017 and 2018. These events are recoverable offline where there is no limit on the processing time except in the cases where timeouts are caused by infinite loops which may require further fixes.
\item \textbf{HLT Error}: If an algorithm aborts processing in an `error' state, the event is sent to the HLTError debug stream. This can occur if, for example, a certain feature cannot be retrieved and sufficient protection is not put in place. These events are recoverable offline, if the cause of the error is fixed.
\item \textbf{PU Crash}: If an algorithm or processing node crashes (due to failure in algorithm reconstruction, an algorithm using too much memory, etc.), the event is sent to the PUCrash debug stream. These events are recoverable offline if the cause of the crash is fixed.
\item \textbf{Missing data}: If some data are missing at the HLT, the event is sent to the HLTMissingData debug stream. This can occur when the L1 result fragment is empty, or if there are inconsistencies in the CTP fragments, or problems recording the EventInfo data, which contains information such as run number, luminosity block number, and bunch crossing identifier. These events are not usually recoverable offline unless fixes can be applied which negate the need for this particular portion of data.
\item \textbf{Possible duplicate}: If the original PU stops responding then the event may get duplicated by being assigned to a second PU and is therefore sent to the possibleDuplicate debug stream. These events are recoverable offline but further checks of data recorded at this point are required.
\item \textbf{Truncated result}: If the HLT result size is larger than a predefined upper limit, the result is truncated and sent to the TruncatedHLTResult stream. These events are usually recoverable offline with the limit on the result size increased.
\item \textbf{Force Accept}: Any problems not covered above. Such events are usually caused by processing problems in a PU due to application crashes where all events being processed by that PU are sent to the ForceAccept debug stream. These events are recoverable offline.
\item \textbf{Late Events}: Events that did not arrive at the SFO to be sent to permanent storage in time but outside the luminosity block boundaries. These events are recoverable offline.
\end{itemize}
 
Table~\ref{tb:debug} gives a breakdown of the number of events in each of the debug streams when the stable beams flag was set and data were recorded for physics during $pp$ data-taking. Over the four years of Run 2, a steady improvement in terms of a reduction in the number of debug stream events can be observed. In addition, the number of events recorded in the physics stream are also listed for each year to put the number of debug stream events in perspective. From this comparison it can be seen that the fraction of events in the debug stream is in principle negligible compared to the overall number of events collected in the physics stream. Out of all events collected for physics analyses, the fractions streamed to the debug stream were
$2.4\times 10^{-4}$, $3.5\times 10^{-5}$, $3.2 \times 10^{-6}$ and $2 \times 10^{-7}$ in 2015, 2016, 2017 and 2018, respectively.
 
\begin{table}
\centering
\caption{A breakdown of the number of events in the different debug streams for each year of $pp$ data-taking during Run 2. In addition, the total number of events recorded in the physics stream is shown.}
\scriptsize
\begin{tabular}{|l|r|r|r|r|}\hline
\multirow{2}{*}{Stream}& \multicolumn{4}{c|}{Number of Events }\\ \cline{2-5}
& 2015 & 2016 & 2017 & 2018\\ \hline \hline
Physics & 1\,694\,555\,330 & 5\,387\,420\,813 & 5\,649\,311\,254& 6\,400\,342\,575 \\ \hline
HLT Timeout & 258\,938 & 23\,468 & 212& 16\\
PU Timeout & 103\,436 &54\,460 & 5759 & 710\\
HLT Error & 21\,302 & 46\,614 & 897 &82\\
PU Crash &16\,830 & 6024 & 2125 &147\\
Missing Data & 1 & 2219 & 0 &0\\
Possible Duplicate & 1 &  42& 4161 &129\\
Truncated HLT Result & 0 &101 &1 &16\\
Force Accept & 11\,110 & 55\,450 & 4838 &397\\
Late Events & 260 & 482 & 204& 10\\  \hline
Total Debug & 411\,878 & 188\,860  & 18\,197 & 1507\\ \hline \hline
Total Debug w.r.t. Physics & $2.4\times 10^{-4}$ &  $3.5\times 10^{-5}$ & $3.2 \times 10^{-6}$ &  $2 \times 10^{-7}$ \\ \hline \hline
Recovered Events& 402\,671 & 187 \,944 & 18\,001 & 1\,455 \\ \hline
Recovered Events w.r.t. Total Debug & 97.8\% & 99.5\% &  98.9\% &  96.5\% \\ \hline
\end{tabular}
\label{tb:debug}
\end{table}
 
While the number of events in the debug stream might be very small, an attempt to recover them is still made.
The recovery is done offline by rerunning the trigger using the same configuration as online, including relevant databases, except the relevant constraints that caused them to be written out to the corresponding debug stream.
In the case of algorithmic errors, the events can be recovered by applying a patch to the software release with a fix for the cause of the failure before rerunning.
Out of the total number of debug stream events, 97.8\%, 99.5\%, 98.9\% and 96.5\% were recovered in 2015, 2016, 2017 and 2018, respectively.
The lower recovery rate in 2018 was due to a small number of events ending up in the debug stream during a special run dedicated to collecting data from the ALFA detectors due to errors in the monitoring code of the ALFA system.
 
In the case of a very high number of debug stream events in a single run (over 100k events), the automatic recovery is not attempted as this typically points to a persistent software issue which requires fixing prior to the recovery attempt.
Events that cannot be recovered offline typically ended up in the debug stream due to significant detector issues deeming the data not good for physics analyses.
If individual LBs contain a large number of unrecoverable events, the corresponding LBs are not considered for use in physics analyses.
 
Events that are recovered offline from the debug streams are available for physics analyses, and analysis teams are advised to include the processing of the debug stream events to check if any event passes their selection and can be included in the analysis.

\pagestyle{plain}

\section{Online monitoring}
\label{sec:OnlineMonitoring}
 
During an ongoing run, the trigger shifter and experts have a variety of tools available to monitor the trigger system.
The monitoring includes that of trigger and stream rates, as well as monitoring of online reconstructed objects to perform an initial assessment of the quality of the data being taken.
Many of these tools make use of the ATLAS Information Service (IS)~\cite{Ref-IS}, which is the backbone of the information sharing between the various online systems.
The data shared through the IS range from simple numbers up to more complex objects such as histograms.
The IS is used to share information between applications in a distributed environment, flowing around the repository which holds the information provided by the applications.
Based on the IS, the Online Histogram Service (OHS) handles the sharing of histograms between the online applications.
Since the HLT is a distributed system over many PUs the histograms need to be added up and handled in a correct way. This is done by an application called the \textit{Gatherer}~\cite{Renkel:2010zz}.
The monitoring information is permanently stored on a run-wise basis for additional cross-checks and analysis on the offline side. The storage of monitoring information is done by the Monitoring Data Archiving (MDA)~\cite{MDA} application, which stores histograms, and P-BEAST~\cite{Avolio:2017rru} which stores all other time-series data. The histograms produced online are stored in ROOT~\cite{root} files and asynchronously transferred to the offline permanent storage.
 
\subsection{Rate monitoring}
\begin{figure}
\centering
\includegraphics[width=0.85\textwidth]{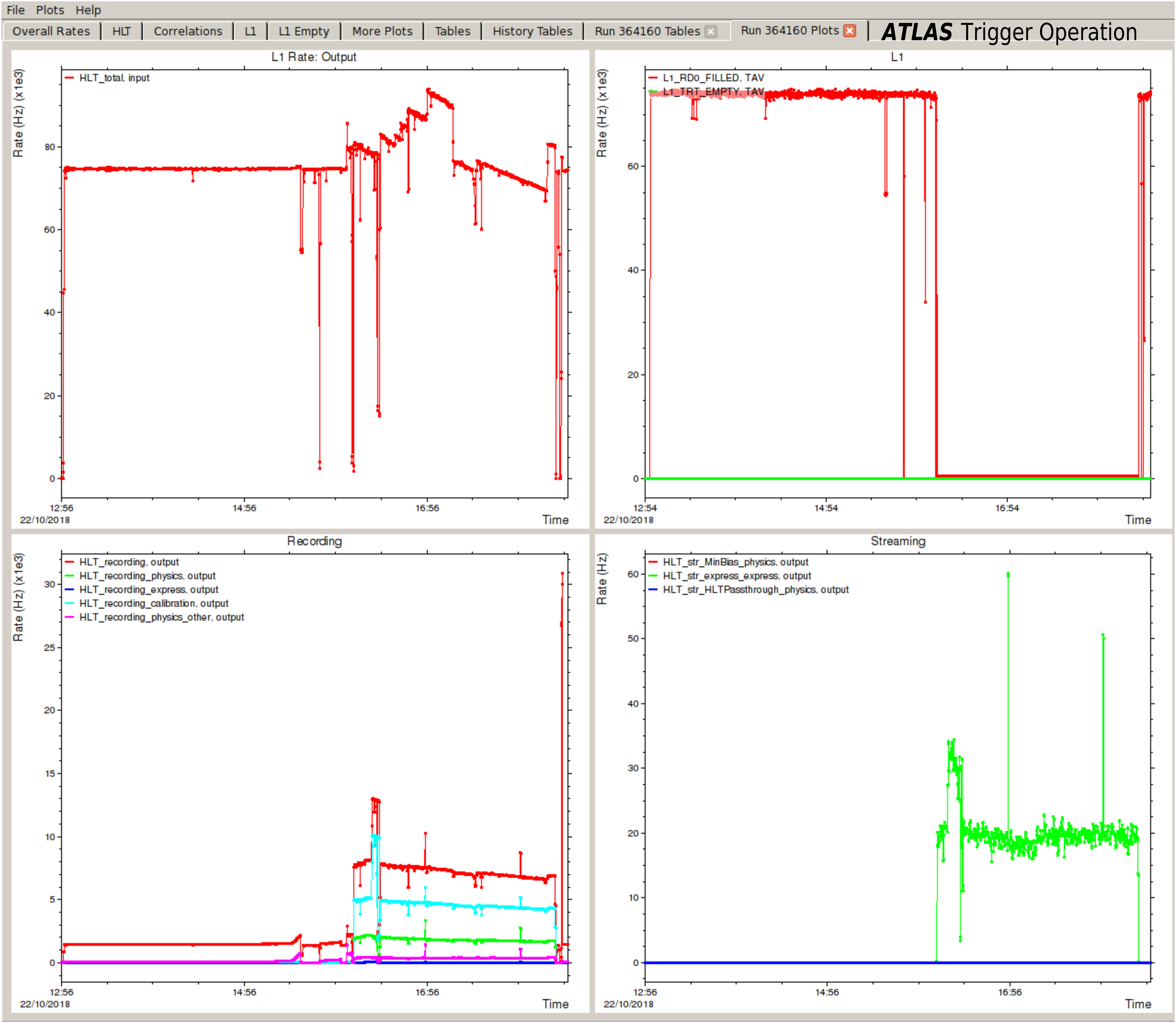}
\caption{The Trigger Rate Presenter is used to monitor the rates of various triggers and streams while taking data.
Shown here is an overview which includes various L1 rates and total HLT output rates, as well as rates in important physics and calibration streams during a data-taking run starting at around 15:35 and ending around 17:40.}
\label{fig:trp}
\end{figure}

Trigger rates are particularly valuable indicators of the trigger performance since they are highly sensitive to detector malfunctions, LHC beam issues, and instantaneous luminosity variations and can therefore often provide a first indication that something is wrong.
Trigger rates are primarily monitored in the control room using the Trigger Rate Presenter (TRP)~\cite{zurNedden:2010wfa}. The TRP is a package that not only displays and monitors the rates, but also archives them for later offline usage.
With the TRP it is possible to monitor the individual rates of all L1 and HLT trigger chains included in the trigger menu.
For L1 items, the rates before and after prescales are available; similarly for HLT chains, the input and output rate before and after prescale can be monitored.
In addition to the rates of individual triggers, the output rates of the different streams are available. An example view of the TRP can be seen in Figure~\ref{fig:trp} displaying four plots of HLT input, output and recording rates, and rates of L1 items and some streams.

\begin{figure}
\centering
\includegraphics[width=1\textwidth]{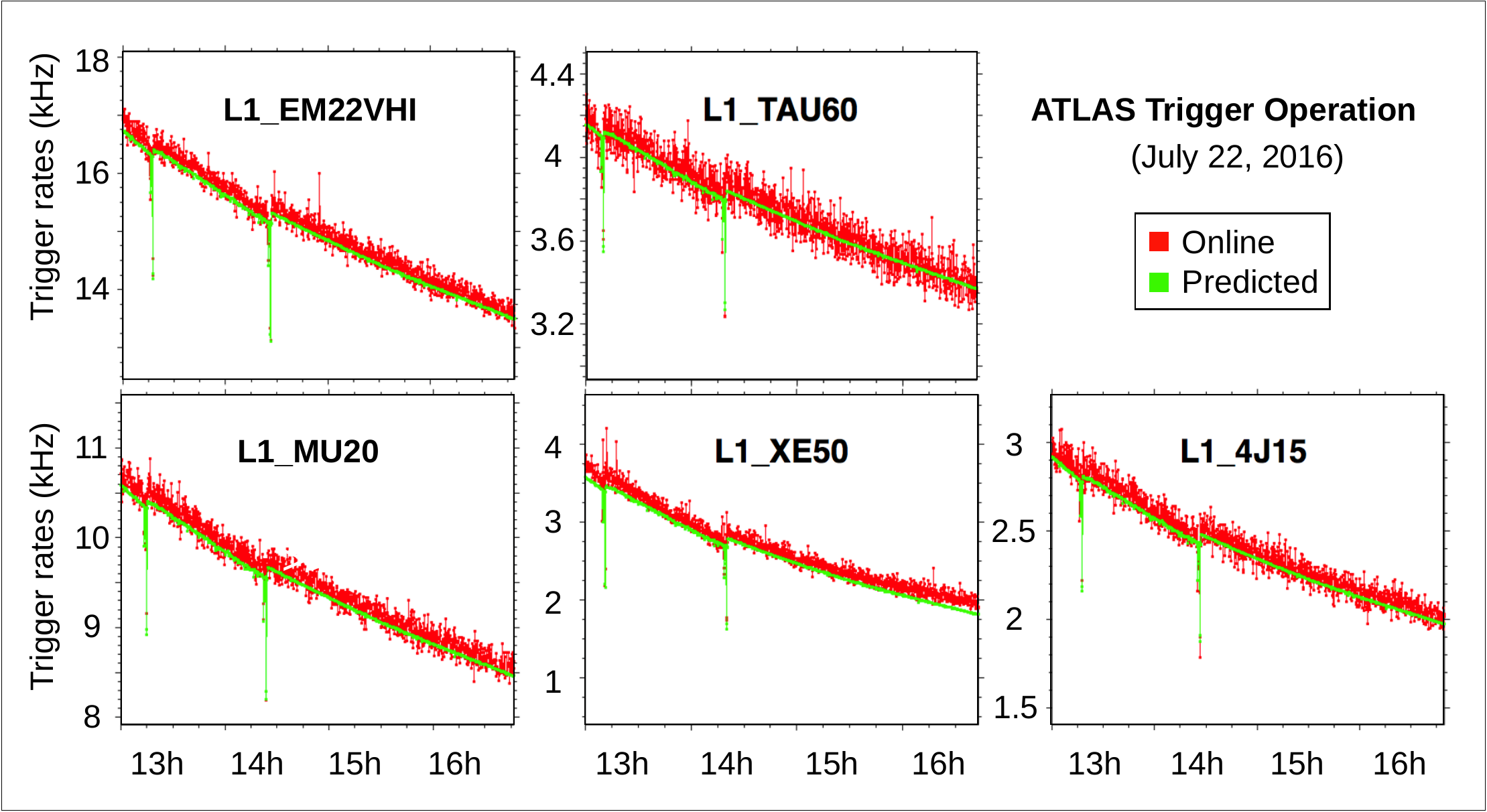}
\caption{Live rates of several L1 trigger items (red) overlaid with the predicted rates (green) as a function of time for one run taken in July 2016. The downward spikes correspond to the luminosity optimisation done by the LHC.}
\label{fig:trpPred}
\end{figure}

In addition, a cross-section / trigger rate monitoring tool (Xmon) was used to compare trigger rates from the ongoing run with rate predictions~\cite{Aukerman:2018pra}.
Xmon provides real-time, luminosity-scaled rate predictions based upon both offline and online trigger cross-section regressions.
The rate predictions are derived offline by fitting past runs with a functional form to the luminosity and pile-up dependence for a given trigger.
This requires the trigger to have been implemented in these past runs but also mitigates issues with predicting rates for triggers which have a potential non-linear rate dependency on luminosity and pile-up.
The predicted rate when running online is then determined using the functional form and the current running parameters (luminosity, trigger prescale, etc.).
The rates of several primary triggers (at L1 and the HLT) with the predictions overlaid can be displayed in the TRP, as shown in Figure~\ref{fig:trpPred}.
Having such predictions for various primary triggers overlaid with the live rates allows shifters to very quickly spot problems with the trigger system or other parts of the detector and proved to be a very useful tool throughout Run 2.

\subsection{Online data quality monitoring}
\begin{figure}
\centering
\includegraphics[width=0.9\textwidth]{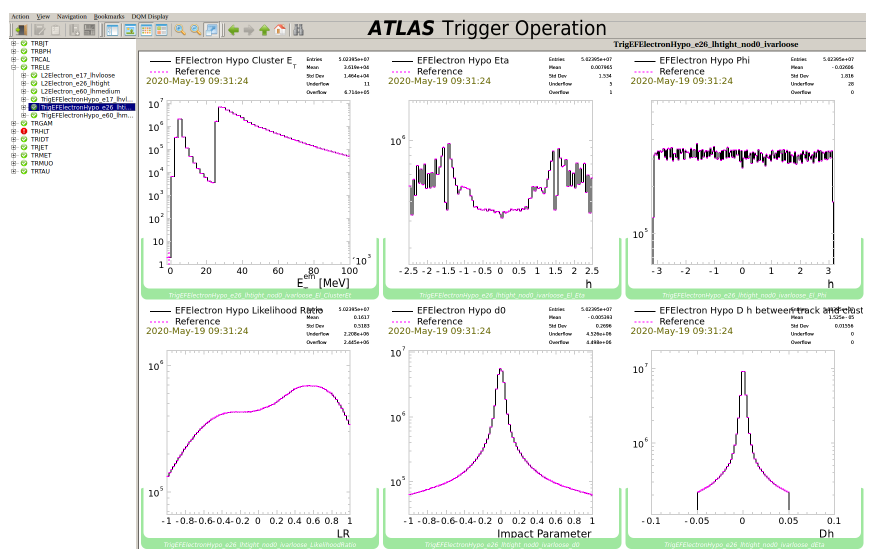}
\caption{The Data Quality Monitoring Display (DQMD) used for the HLT.
The data (black) are compared with the reference (purple) using a Kolmogorov--Smirnov test to compare the shapes of the distributions.
Based on the output of the comparison test, the histograms are flagged either green (good agreement with the reference), yellow (tolerable disagreement with the reference), or red (major disagreement with the reference).}
\label{fig:dqm}
\end{figure}

In addition to the trigger rates, histograms of other quantities from the trigger system are monitored~\cite{Kolos:2649758}.
These include kinematic properties of various objects reconstructed by the trigger, as well as more general infrastructure-related information (e.g.~number of processing slots available in the HLT farm, the status of the magnets, etc.).
The histograms are displayed using the Data Quality Monitoring Display (DQMD)~\cite{Ilchenko_2010}.
The display is organised in a tree structure with histograms from different trigger signatures grouped together.
The histograms are displayed along with a reference histogram and are flagged with different colours depending on the outcome of automatic tests.
These tests can be as simple as checking that a histogram is filled, or can be more complicated such as checking the level of agreement using a Kolmogorov--Smirnov test between the live histogram and its reference (and flagging the histogram as green for good, yellow for acceptable, red for bad agreement).
Other types of quality assessment algorithms may be also executed for the histograms independently of the reference, such as comparing the mean value of a distribution with a certain threshold.
The coloured flag is propagated up the tree structure, so even if only one histogram is flagged red, it is easily spotted from the overall view.
The histogram references are taken from a good run with similar running conditions, and are updated periodically as running conditions change.
An example of the data quality display showing several histograms of the electron candidates reconstructed in the HLT is shown in Figure~\ref{fig:dqm}.
 
A second GUI called the Online Histogram Presenter (OHP)~\cite{OHP} is available to display histograms.
The OHP display does not include the colour-coded flags, but is used to display additional, more detailed histograms that are used by experts to debug problems.
It also has an easy-to-use web display that allows the user to interact with the histograms (zoom in, etc.).
 
\subsection{P-BEAST and shifter assistant}
 
\begin{figure}
\centering
\includegraphics[width=1\textwidth]{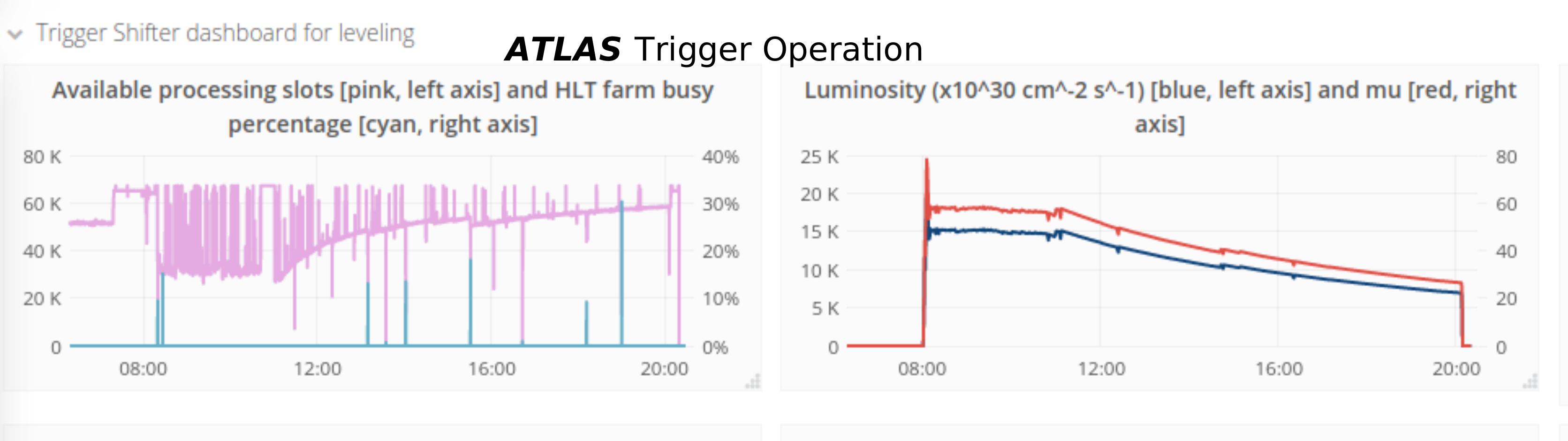}
\includegraphics[width=1\textwidth]{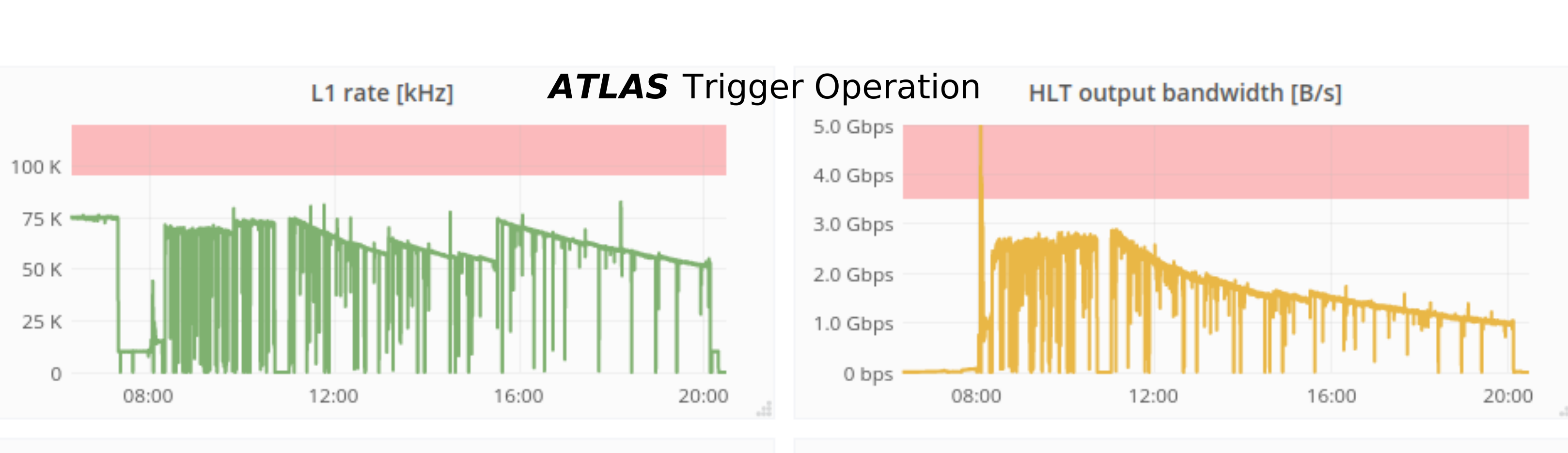}
\caption{An example of a P-BEAST dashboard as used by the shifters during 2017 operation. The HLT farm availability (top left), the luminosity and the pile-up,`mu' (top right), the total L1 accept rate (bottom left), and the total HLT output bandwidth (bottom right) are shown.}
\label{fig:pbeast}
\end{figure}

P-BEAST is a web-based monitoring tool used by shifters and experts in addition to the other available tools to monitor the system and debug problems.
P-BEAST displays data from the IS graphically using Grafana~\cite{grafana} dashboards.
The dashboard developed for the trigger displays, as a function of time, rates and bandwidths for various streams, the prescale keys in use, the number of free processing slots in the HLT farm, memory usage, read-out request rates, luminosity, dead times from detectors, etc.
It provides a very quick overview of the current state of the running system, as well as the history.
An example dashboard is shown in Figure~\ref{fig:pbeast}.
This dashboard was heavily used during 2017 data-taking with the 8b4e filling scheme where luminosity levelling was required in order to stay within the trigger resource limitations, and provided the shifters an easy overview to check that the limits were respected.
 
To help shifters catch problems, an alert system called the Shifter Assistant~\cite{Kazarov:2648925} is used in ATLAS to flag certain possible issues.
The alerts are displayed in a webpage, with a pop-up notification for new alerts.
In the trigger, these alerts notify shifters if rates of different streams drop below or rise above certain specified values (depending on the stream).
It also notifies shifters of plots that are flagged red in the DQMD, and sends periodic reminders to the shifters to check rates and histograms.
 
\subsection{Data-taking anomalies}
While precaution is taken to prevent anomalies in data-taking from occurring, quick action needs to be taken in case smooth operation is not ensured any more. In such a situation, the trigger shifter in the control room and the on-call experts work together to ameliorate the issue.
The time taken to solve a given issue can vary widely, and depends on the severity of the problem, and the subsystems involved.
All the online tools described above are meant to assist and provide as much information as possible to quickly identify the source of the issue. 
 
A change in rate is one of the most common problems where trigger rates can be either too high or too low.
If rates are high enough to cause issues with the data flow, the cause needs to be identified and fixed promptly as it prevents data from being recorded. The circumstance in which the change in rate occurred is important to understand, and below are some examples of information that can help identify a given problem:
\begin{itemize}
\item The time of occurrence, i.e.\ in the middle of a data-taking run or at the beginning of the run.
\item Any changes made prior to the observation of the high rate in the trigger system or sub-detector systems, e.g.\ a new software release, change of prescales, etc.
\item The current status of sub-detector systems and if they are in good condition. Good communication between different members of the shift crew in the control room is essential to efficiently forward information to the experts.
\item Changes from the LHC, e.g.\ luminosity optimisation during a run.
\end{itemize}
With help of the TRP, the rates of each stream or individual L1 item or HLT chain can be displayed.
Once the type of trigger has been identified, the underlying problem can be further narrowed down.
It is possible that a wrong prescale key or an incorrect combination of L1 and HLT prescale keys together with an incorrect bunch group configuration was loaded during the run.
Noisy cells from sub-detectors (e.g.\ in the calorimeter) can lead to spikes in trigger rates, which can be identified using the DQMD to check the positions of reconstructed objects.
No or significantly lower rate in e.g.\ a group of trigger chains typically points to a problem with a sub-detector system or incorrect algorithm configuration in the HLT software.
For example, if some stations of the muon detectors are not working, lower rates in the muon trigger chains can be observed, or similarly, a lower rate in $b$-jet and electron chains may indicate a problem with the data quality of the inner detector.
With the help of the trigger rate predictions, changes in the rates during a run or right from the beginning can be spotted quickly.
 
High dead time, and consequently severe problems with the data flow, can have several causes and often requires a combined investigation by trigger and DAQ experts.
Since none of the DAQ/HLT components~\cite{DAQpaper} can generate a busy signal (unlike the front-end systems) they need to rely on backpressure between components to temporarily stop the data flow.
As the buffer space fills up, the back-pressure builds up until the data flow is suspended completely.
Backpressure can arise in the HLT if there are algorithms running in the HLT that require too much CPU time for the HLT farm to be able to process events quickly enough, so that no processing units in the HLT farm remain available to process new events.
In this case, chains running CPU-intensive algorithms are typically prescaled or disabled in order to free up CPU resources until the CPU usage of the algorithms can be reduced.
The HLT can also cause back-pressure if it tries to write out too much data.
This can happen if there is a very high rate of events being written into a debug stream (if, for example, an algorithm is exiting in an error state in every event it runs).
Depending on which stream they are written to, this could indicate a detector problem, or a problem with a specific chain or algorithm running in the HLT.
In the latter cases, the usual intermediate solution is to disable or prescale the offending chains until a more permanent solution can be implemented.
 
The main operational issues encountered during Run 2 that caused data losses were typically too high rates, induced either by changes in the LHC conditions (e.g.\ increase in bunch intensities, different filling schemes, luminosity optimisation during a run) or sub-detector problems. 
 
\pagestyle{plain}
 
\section{Offline monitoring and data quality assessment}
\label{sec:OfflineMonitoringDQ}
 
In addition to monitoring the data online as it is being collected (Section~\ref{sec:OnlineMonitoring}), an assessment of the quality of the data is carried out offline after the data are fully reconstructed.
This allows more detailed studies of the trigger efficiency relative to offline objects.
The data quality (DQ) assessment procedures for all of ATLAS are described in Ref.~\cite{Aad:2019hmz}.
 
\begin{figure}
\centering
\includegraphics[width=\textwidth]{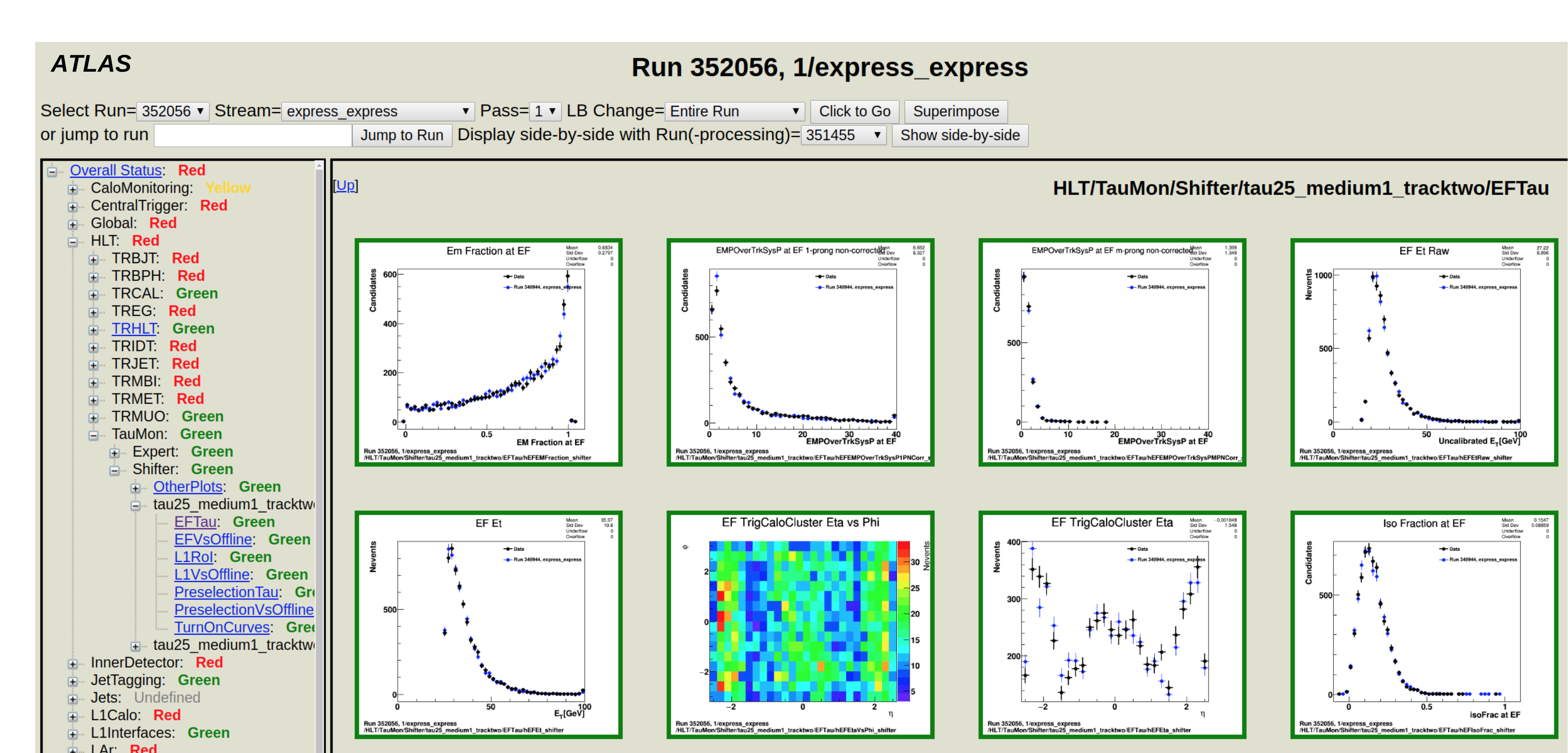}
\caption{An example of the web-display for the offline DQ assessment.
Shown here are various plots for hadronically decaying $\tau$-lepton candidates as reconstructed in the HLT.
The data (black points) are compared with a reference (blue points), and flagged as green, yellow, or red depending on how well they agree, according to predefined algorithms.
The reference run is updated periodically to reflect significant condition changes, such as updated trigger menus.
This example shows run 352\,056,which was recorded in June 2018.
Error bars include the statistical uncertainties only.
}
\label{fig:offlinedqm}
\end{figure}

The express stream is the main data stream used for the offline DQ assessment, and this stream is fully reconstructed within a day after data are recorded.
Signature on-call experts (see Section~\ref{sec:shifter}) assess the performance of the trigger chains by looking at a series of histograms which are chosen to sample all relevant aspects of the particular trigger signature.
The histograms can include kinematic distributions, efficiencies, resolutions, and comparisons between trigger objects and the corresponding offline reconstructed objects.
As shown in Figure~\ref{fig:offlinedqm}, the histograms are arranged in a web-based display that includes a reference histogram, and a colour-coded flag based on how well the histogram and its reference agree. Mismatches (yellow in case of slight deviations, or red) can be due to either actual problems, outdated references (e.g.\ in the luminosity ramp-up phase at the start of a data-taking period), or low sample size and need to be followed up and understood.
In some cases, the number of events collected in the express stream is not sufficient to perform a full assessment of the DQ.
In these cases, the express stream is used as a preliminary check, and the final assessment is done when looking at the same histograms but produced by processing all events collected in the physics stream. These histograms are typically available a few days after the express stream is processed.
 
If there are problems spotted during the assessment, a so-called data quality \emph{defect} is set and recorded in the data quality database.
A defect denotes an LB (or several LBs up to the entire run) in which the data-acquisition was not optimal.
Many different defects are defined, so that each type of problem can be tracked individually.
They can be either tolerable or intolerable.
Tolerable defects are used to flag slight abnormalities in the data that do not have a large impact on physics (e.g.~spikes in the calorimeter energy, small loss of coverage from some detector, etc.).
Intolerable defects are used to flag data that should not be used for physics (e.g.~large loss of coverage from some detector, incorrect trigger configuration, etc.).
 
The DQ and debug stream on-call expert is responsible for collecting the assessment outcome from the L1 systems and the trigger signatures and reporting them to the ATLAS DQ group. Any relevant findings from the other ATLAS subsystems are also reported back to the trigger community.
 
The collection of all defects from the DQ assessments of the trigger system as well as of all other parts of the ATLAS detector constitute the basis for the creation of the Good Runs List (GRL).
The GRL is the list of all LBs of all runs collected by ATLAS which are deemed good for physics analysis.
Generally, any LB which has been labelled by any intolerable defect is vetoed from the GRL, as described in Ref.~\cite{Aad:2019hmz}.
 
A summary of the DQ efficiency for the trigger system for each year together with the overall DQ efficiency for the ATLAS experiment is given in Table~\ref{DQsummary}, which is extracted from Ref.~\cite{Aad:2019hmz}.
The DQ efficiency is calculated relative to the recorded integrated luminosity.
It accounts for any data that are rejected on the grounds of data quality and therefore does not take into account the events that were not recorded due to detector-related problems.
The DQ efficiency is presented in terms of a luminosity-weighted fraction of good quality data recorded during stable beams periods.
Only periods during which the recorded data were intended to be used for physics analysis are considered for the DQ efficiency.
The trigger-specific and total ATLAS DQ efficiency both improved over the course of \runii.
Over the full \runii, the trigger and ATLAS DQ efficiencies are 99.6\% and 95.6\%, respectively, for the standard $\sqrt{s}=13$~TeV $pp$ dataset, with a corresponding 139 fb$^{-1}$ of data being good for physics analysis.
The inefficiencies of the trigger system where the data were marked with defects as not good for ATLAS are mainly due to either hardware problems (read-out from detectors due to problems further upstream) in the L1 system or an incorrect trigger configuration being loaded.
For L1 it was mostly due to detector coverage and a problem with the CTP cables.
The data loss from events that were not recorded due to problems in the TDAQ system is very small and was reduced during each successive \runii data-taking year. It accounts for a data-taking efficiency loss of 3.16\% in 2015, 2.22\% in 2016, 1.04\% in 2017 and 0.671\% in 2018.
 
\begin{table}[hb]
\begin{center}
\caption{Luminosity-weighted relative detector uptime and good DQ efficiencies (in \%) during stable beams $pp$ collision physics runs at $\sqrt{s}=13$~\TeV\ for 2015 (July to November), 2016 (April to October), 2017 (June to November) and 2018 (April to October) for the trigger system (L1 and HLT) and overall for the ATLAS detector. Numbers were extracted from Ref.~\cite{Aad:2019hmz}.
}
\begin{tabular}{|l||l|c|c|c|c|}
\hline
\multirow{2}{*}{Year} & \multirow{2}{*}{Dataset} &\multicolumn{2}{c|}{Trigger DQ Eff.} & \multirow{1}{*}{ATLAS DQ Eff.} & \multirow{1}{*}{Integrated Luminosity}\\ \cline{3-4}
&  &\multicolumn{1}{c|}{L1  [\%]}         &     \multicolumn{1}{c|}{HLT  [\%]} &  [\%] & of good quality data \\
\hline
\multirow{2}{*}{2015} & $pp$ @ 13~\TeV\ (50~ns)  &   100.00~~     &  99.94 & 88.77& 84 pb$^{-1}$\\
& $pp$ @ 13~\TeV                &   99.97  &  99.76 & 88.79& 3.2 fb$^{-1}$~~\\
\hline
2016 &  $pp$ @ 13~\TeV  & 98.33 & 100.00~~ & 93.07 & 33 fb$^{-1}$\\
\hline
2017 &  $pp$ @ 13~\TeV  & 99.95 & 99.96 & 95.67&44 fb$^{-1}$\\
\hline
2018 &  $pp$ @ 13~\TeV  & 99.99 & 99.99 & 97.46 & 59 fb$^{-1}$\\
\hline
2015--2018 & $pp$ @ 13~\TeV & 99.57 & 99.94 & 95.60 & 139 fb$^{-1}$~~~\\
\hline
\end{tabular}
 
\label{DQsummary}
\end{center}
\end{table}

 
\FloatBarrier
\section{\label{sec:conclusion}Conclusion}
 
The operation of the ATLAS trigger system during \runii of the LHC was highly successful for both $pp$ and heavy-ion data-taking as well as for a variety of special runs. The recorded data are of high quality with only minimal losses due to operational problems.
The trigger's excellent performance and flexibility allowed collection of $139\,\mathrm{fb}^{-1}$ of high-quality $pp$ collision data at $\sqrt{s}=13$~\TeV\ over a wide range of running conditions, with peak instantaneous luminosities ranging from $0.5\times\lumi{e34}$ to $2.1\times\lumi{e34}$ and 28--60 pile-up collisions.
A variety of tools and procedures, as detailed in this paper, have been used in both the offline and online environments to efficiently validate, monitor and assess the operation and the performance of the trigger system.
They allowed maximal exploitation of the collisions delivered by the LHC in \runii for physics analyses, while respecting the limitations of the trigger system in terms of maximum read-out rates, available HLT computing resources and sustainable storage output bandwidth.
 
\section*{Acknowledgements}
 

We thank CERN for the very successful operation of the LHC, as well as the
support staff from our institutions without whom ATLAS could not be
operated efficiently.
 
We acknowledge the support of ANPCyT, Argentina; YerPhI, Armenia; ARC, Australia; BMWFW and FWF, Austria; ANAS, Azerbaijan; SSTC, Belarus; CNPq and FAPESP, Brazil; NSERC, NRC and CFI, Canada; CERN; CONICYT, Chile; CAS, MOST and NSFC, China; COLCIENCIAS, Colombia; MSMT CR, MPO CR and VSC CR, Czech Republic; DNRF and DNSRC, Denmark; IN2P3-CNRS and CEA-DRF/IRFU, France; SRNSFG, Georgia; BMBF, HGF and MPG, Germany; GSRT, Greece; RGC and Hong Kong SAR, China; ISF and Benoziyo Center, Israel; INFN, Italy; MEXT and JSPS, Japan; CNRST, Morocco; NWO, Netherlands; RCN, Norway; MNiSW and NCN, Poland; FCT, Portugal; MNE/IFA, Romania; MES of Russia and NRC KI, Russia Federation; JINR; MESTD, Serbia; MSSR, Slovakia; ARRS and MIZ\v{S}, Slovenia; DST/NRF, South Africa; MINECO, Spain; SRC and Wallenberg Foundation, Sweden; SERI, SNSF and Cantons of Bern and Geneva, Switzerland; MOST, Taiwan; TAEK, Turkey; STFC, United Kingdom; DOE and NSF, United States of America. In addition, individual groups and members have received support from BCKDF, CANARIE, Compute Canada and CRC, Canada; ERC, ERDF, Horizon 2020, Marie Sk{\l}odowska-Curie Actions and COST, European Union; Investissements d'Avenir Labex, Investissements d'Avenir Idex and ANR, France; DFG and AvH Foundation, Germany; Herakleitos, Thales and Aristeia programmes co-financed by EU-ESF and the Greek NSRF, Greece; BSF-NSF and GIF, Israel; CERCA Programme Generalitat de Catalunya and PROMETEO Programme Generalitat Valenciana, Spain; G\"{o}ran Gustafssons Stiftelse, Sweden; The Royal Society and Leverhulme Trust, United Kingdom.
 
The crucial computing support from all WLCG partners is acknowledged gratefully, in particular from CERN, the ATLAS Tier-1 facilities at TRIUMF (Canada), NDGF (Denmark, Norway, Sweden), CC-IN2P3 (France), KIT/GridKA (Germany), INFN-CNAF (Italy), NL-T1 (Netherlands), PIC (Spain), ASGC (Taiwan), RAL (UK) and BNL (USA), the Tier-2 facilities worldwide and large non-WLCG resource providers. Major contributors of computing resources are listed in Ref.~\cite{ATL-SOFT-PUB-2020-001}.
 

\printbibliography

\clearpage \input{atlas_authlist}

\end{document}

%% file: atlas_authlist.tex
 
\begin{flushleft}
\hypersetup{urlcolor=black}
{\Large The ATLAS Collaboration}

\bigskip

\AtlasOrcid[0000-0002-6665-4934]{G.~Aad}$^\textrm{\scriptsize 102}$,    
\AtlasOrcid[0000-0002-5888-2734]{B.~Abbott}$^\textrm{\scriptsize 128}$,    
\AtlasOrcid{D.C.~Abbott}$^\textrm{\scriptsize 103}$,    
\AtlasOrcid[0000-0002-2788-3822]{A.~Abed~Abud}$^\textrm{\scriptsize 36}$,    
\AtlasOrcid[0000-0002-1002-1652]{K.~Abeling}$^\textrm{\scriptsize 53}$,    
\AtlasOrcid[0000-0002-2987-4006]{D.K.~Abhayasinghe}$^\textrm{\scriptsize 94}$,    
\AtlasOrcid[0000-0002-8496-9294]{S.H.~Abidi}$^\textrm{\scriptsize 166}$,    
\AtlasOrcid[0000-0002-8279-9324]{O.S.~AbouZeid}$^\textrm{\scriptsize 40}$,    
\AtlasOrcid{N.L.~Abraham}$^\textrm{\scriptsize 155}$,    
\AtlasOrcid[0000-0001-5329-6640]{H.~Abramowicz}$^\textrm{\scriptsize 160}$
% ...
\bigskip
\\

% ...
$^{36}$CERN, Geneva; Switzerland.\\
% ...
$^{40}$Niels Bohr Institute, University of Copenhagen, Copenhagen; Denmark.\\
% ...
$^{53}$II. Physikalisches Institut, Georg-August-Universit\"{a}t G\"ottingen, G\"ottingen; Germany.\\
% ...
$^{94}$Department of Physics, Royal Holloway University of London, Egham; United Kingdom.\\
% ...
$^{102}$CPPM, Aix-Marseille Universit\'e, CNRS/IN2P3, Marseille; France.\\
$^{103}$Department of Physics, University of Massachusetts, Amherst MA; United States of America.\\
% ...
$^{128}$Homer L. Dodge Department of Physics and Astronomy, University of Oklahoma, Norman OK; United States of America.\\
% ...
$^{155}$Department of Physics and Astronomy, University of Sussex, Brighton; United Kingdom.\\
% ...
$^{160}$Raymond and Beverly Sackler School of Physics and Astronomy, Tel Aviv University, Tel Aviv; Israel.\\
% ...
$^{166}$Department of Physics, University of Toronto, Toronto ON; Canada.\\
% ...

\end{flushleft}
